\documentclass[a4paper,fleqn,usenatbib]{mnras}
\usepackage{graphicx,xspace}
\usepackage{dcolumn, xcolor}
\usepackage{bm}
\usepackage[fleqn]{amsmath}
\usepackage{amsfonts,amssymb} 
\usepackage[e]{esvect}
\usepackage{hyperref}
\usepackage{bigints}
\usepackage{xparse}
\usepackage{rotating}
\usepackage[utf8]{inputenc}
\usepackage[T1]{fontenc}
\usepackage{mathptmx}
\usepackage{subfigure}
\usepackage{ wasysym }
\usepackage{twoopt}
\usepackage[T1]{fontenc}
\usepackage{ae,aecompl,fancyvrb}

\newcommand{\tr}[1]{\ensuremath{\mathrm{tr}(\tensor{#1})}}
\newcommand{\KU}{Kummer-$U$ function\xspace}

\newcommand{\KUT}{Kummer-$U$ or Tricomi function\xspace}

\newcommand{\DL}{Debye length\xspace}

\providecommand\const{\ensuremath{\mathrm{const.}}}
\NewDocumentCommand\hyp{oo}{
  \IfNoValueTF{#1}{\ensuremath{\ \mathrm{_{[]}F_{[]}}}}
     {\IfNoValueTF{#2}{\mbox{\ensuremath{\ _{[#1]}\mathrm{F}_{[]}}}}
       {\mbox{\ensuremath{\ _{[#1]}\mathrm{F}_{[#2]}}}}
}}

\NewDocumentCommand\Uka{oo}{
  \IfNoValueTF{#1}{\ensuremath{\ {\matcal{U}^{[#2]}}}}
     {\IfNoValueTF{#2}{\mbox{\ensuremath{\ ^{[#1]}\mathcal{U}}}}
       {\mbox{\ensuremath{\ ^{[#1]}\mathcal{U}^{[#2]}}}}
     }
   }
\NewDocumentCommand\uka{oo}{
  \IfNoValueTF{#1}{\ensuremath{\ {\mathrm{^{[]}u}}}}
     {\IfNoValueTF{#2}{\mbox{\ensuremath{\ ^{[#1]}\mathrm{u}}}}
       {\mbox{\ensuremath{\ ^{[#1]}\mathrm{u}}}}
}}
\NewDocumentCommand\Wka{oo}{
  \IfNoValueTF{#1}{\mbox{\ensuremath{\ \ ^{{[]}}\mathcal{W}^{[]}}}}
     {\IfNoValueTF{#2}{\mbox{\ensuremath{\ ^{[#1]}\mathcal{W}^{[]}}}}
       {\mbox{\ensuremath{\ ^{[#1]}\mathcal{W}^{[#2]}}}}
}}
\NewDocumentCommand\wka{oo}{
  \IfNoValueTF{#1}{\ensuremath{\ \mathrm{^{[]}w}}}
     {\IfNoValueTF{#2}{\mbox{\ensuremath{\ ^{[#1]}\mathrm{w}}}}
       {\mbox{\ensuremath{\ ^{[#1]}\mathrm{w}}}}
}}
\NewDocumentCommand\Vka{ooo}{
  \IfNoValueTF{#1}{\mbox{\ensuremath{\ \ ^{{[]}}\mathcal{V}^{[]}^{#3}}}}
     {\IfNoValueTF{#2}{\mbox{\ensuremath{\ ^{[#1]}\mathcal{V}^{[]}^{#3}}}}
       {\mbox{\ensuremath{\ ^{[#1]}\mathcal{V}^{[#2]}^{#3}}}}
}}\NewDocumentCommand\sinf{oo}{
  \IfNoValueTF{#1}{\ensuremath{\sin\varphi}}
     {\IfNoValueTF{#2}{\ensuremath{\sin^{#1}\varphi}}
       {\ensuremath{\sin^{#1}#2}}
}}
\NewDocumentCommand\cosf{oo}{
  \IfNoValueTF{#1}{\ensuremath{\cos\varphi}}
     {\IfNoValueTF{#2}{\ensuremath{\cos^{#1}\varphi}}
       {\ensuremath{\cos^{#1}#2}}
}}
\NewDocumentCommand\sint{oo}{
  \IfNoValueTF{#1}{\ensuremath{\sin\vartheta}}
     {\IfNoValueTF{#2}{\ensuremath{\sin^{#1}\vartheta}}
       {\ensuremath{\sin^{#1}#2}}
}}
\NewDocumentCommand\cost{oo}{
  \IfNoValueTF{#1}{\ensuremath{\cos\vartheta}}
     {\IfNoValueTF{#2}{\ensuremath{\cos^{#1}\vartheta}}
       {\ensuremath{\cos^{#1}#2}}
}}

\newcommand{\dt}{\ensuremath{\mathrm dt}}
\newcommand{\dtheta}{\ensuremath{\mathrm d}\vartheta}
\newcommand{\dphi}{\ensuremath{\mathrm d}\varphi}

\providecommand\tensor[1]{\ensuremath{\overleftrightarrow{#1}}}

\newcommand{\GA}[1]{\ensuremath{\Gamma\left(#1\right)}}

\newcommand{\md}{\ensuremath{\mathrm{d}}}
\providecommand\half{\ensuremath{\frac{1}{2}}}
\providecommand\thalf{\ensuremath{\frac{3}{2}}}

\providecommand\aap{Astron. \& Astrophys.}

\providecommand\apj{Astrophs. J.}

\newcolumntype{R}{>{$}r<{$\ \ =}}
\newcolumntype{C}{>{$}c<{$\hspace{0.cm}}}
\newcolumntype{K}{>{$}r<{$\hspace{0.cm}}}
\newcolumntype{L}{>{\hspace{-0.1cm}$}l<{$}}
\newcolumntype{X}{>{\ $}l<{$}}

\let\oldsqrt\sqrt
\def\sqrt{\mathpalette\DHLhksqrt}
\def\DHLhksqrt#1#2{%
\setbox0=\hbox{$#1\oldsqrt{#2\,}$}\dimen0=\ht0
\advance\dimen0-0.2\ht0
\setbox2=\hbox{\vrule height\ht0 depth -\dimen0}%
{\box0\lower0.4pt\box2}}

\allowdisplaybreaks

\title{The $\kappa$-cookbook: a novel generalizing
approach to unify $\kappa$-like distributions for plasma particle modeling}

\author[K. Scherer et al.]{K. Scherer,$^{1,2}$\thanks{kls@was.tp4.rub.de}
  and E. Husidic$^{1,3}$, M. Lazar,$^{1,3}$, H. Fichtner,$^{1,2}$
\\
$^{1}$Institut f\"ur Theoretische Physik, Lehrstuhl IV:
  Plasma-Astroteilchenphysik, Ruhr-Universit\"at Bochum, D-44780 Bochum,
  Germany\\
$^{2}$Research Department, Plasmas with Complex Interactions,
  Ruhr-Universit\"at Bochum, 44780 Bochum, Germany \\
$^{3}$Centre for Mathematical Plasma Astrophysics, 
      3001 Leuven Belgium
}
\pubyear{}
\date{}
\VerbatimFootnotes
\begin{document}
\label{firstpage}
\pagerange{\pageref{firstpage}--\pageref{lastpage}}
\maketitle

\begin{abstract}
  In the literature different so-called $\kappa$-distribution
  functions are discussed to fit and model the velocity (or energy)
  distributions of solar wind species, pickup ions or magnetospheric
  particles.  Here we introduce a generalized (isotropic)
  $\kappa$-distribution
  as a "cookbook", which admits as special cases, or "recipes", all
  the other known versions of $\kappa$-models.
  A detailed analysis of the generalized distribution function is
  performed, providing general analytical expressions for the velocity
  moments, \DL, and entropy, and pointing out a series of general
  requirements that plasma distribution functions should satisfy. From
  a contrasting analysis of the recipes found in the literature, we
  show that all of them lead to almost the same macroscopic parameters
  with a small standard deviation between them. However, one of these recipes called 
  the regularized $\kappa$-distribution provides a functional alternative for macroscopic 
  parameterization without any constraint for the power-law  exponent $\kappa$.
\end{abstract}

\begin{keywords}plasmas,  Sun: heliosphere, solar wind, methods: data analysis 
  \end{keywords}
\section{\label{sec:1}Introduction}

Different so-called Kappa- or $\kappa$-distributions are widely applied in 
    space physics to model the suprathermal tails of particle energy or velocity 
    distributions in collision-poor and dilute astrophysical plasma environments. 
    While the core of such distributions can be well fitted by a Maxwellian, the enhanced 
    wings of the distribution are best approximated by power-laws \citep{Pierrard-Lazar-2010}.
    The original $\kappa$-distribution has been defined over 50 years ago in a rather 
    ad-hoc manner by \cite{Olbert-1968} and \cite{Vasyliunas-1968} to reproduce the velocity 
    distributions of magnetospheric electrons. Since then various attempts have been made 
    to derive the $\kappa$-distribution theoretically in a more rigorous 
    way in prescribed plasma setups, e.g., \cite{Hasegawa-etal-1985}, who derived a Kappa-like energy 
    distribution for a plasma in a superthermal radiation field, \cite{Ma-Summers-1998}, who 
    found a $\kappa-$distribution to be the solution of the Fokker-Planck equation with the 
    inclusion of stationary whistler turbulence, and \cite{Yoon-2014}, who self-consistently solved the 
    problem of an isotropic electron distribution that is in equilibrium with the electrostatic 
    Langmuir turbulence and found a $\kappa$-distribution with specifically $\kappa = 9/4$ (see
    below for an explanation of the $\kappa$-parameter). 
    Further effort to put the $\kappa$-distributions on a more solid theoretical ground resulted 
    in a generalization of the standard Boltzmann-Gibbs entropy by \cite{Tsallis-1988} and 
    \cite{Treumann-Jaroschek-2008} in order to account for non-equilibrium distributions and 
    a nonadditive entropy
    \citep[see also][]{Fichtner-etal-2018}.
    
Besides its original employment by \cite{Olbert-1968} and \cite{Vasyliunas-1968} to describe electrons
    in Earth's magnetosphere, the family of $\kappa$-distributions finds practical application also in 
    many other areas of space physics, e.g., in the study of the interplanetary medium and planetary
    magnetospheres \citep{Maksimovic-etal-1997, Maksimovic-etal-2005, Pierrard-Lazar-2010}, the outer 
    heliosphere \citep{Zank-etal-2010,Fahr-etal-2016, Fahr-etal-2017,Heerikhuisen-etal-2019}, especially for charge exchange processes \citep{Heerikhuisen-etal-2015} and to fit IBEX observations of neutral atoms \citep{Desai-etal-2012}, the interstellar medium \citep{Davelaar-etal-2018} 
    and the intergalactic medium \citep{deAvillez-etal-2018}. Recently the $\kappa$-distributions
    found their way even into experimental physics \citep{Webb-etal-2012, Elkamash-Kourakis-2016}.

Following the terminology in \cite{Scherer-etal-2017d, Scherer-etal-2019b} we will call the 
    original version by \cite{Vasyliunas-1968} the standard $\kappa$-distribution (SKD), defined as 
    \begin{align}\label{eq:skd}
    f_{SKD}(\vec{r},\vec{v},t) =
    \frac{n(\vec{r},t)}{\sqrt{\pi^{3}\kappa}\Theta^{3}}
    \frac{\GA{\kappa}}{\GA{\kappa-\half}}
    \left[1 + \frac{v^{2}}{\kappa\Theta^{2}}\right]^{-(\kappa+1)}\,,
    \end{align}
    where $\Gamma(\mu)$ is the (complete) Gamma function of argument $\mu$, $\Theta$ is 
    defined as the most probable speed, which normalizes the particle velocity $\vec{v}$
    and its magnitude $v$, respectively, and $n(\vec{r},t)$ is the number 
    density, which in general can depend on location $\vec{r}$ and time $t$. The $\kappa$-parameter is
    a free parameter and serves as a measure of the departure of the SKD from its 
    Maxwellian core \citep{Vasyliunas-1968} and thus describes the high-enrgy power-law tails of the distribution. Equation \eqref{eq:skd} and all the following specified distributions 
    are normalized to number density, and for $\kappa \to \infty$ the SKD approaches its Maxwellian core. 
    Beside the SKD many other $\kappa$-like versions have been proposed in the literature 
    \citep[e.g,.][]{Yoon-2012, Livadiotis-McComas-2013, Lazar-etal-2015, Treumann-Baumjohann-2014,Lazar-etal-2017, Scherer-etal-2017}, 
    of which most suffer under one or more deficiencies listed in the next paragraph. Some authors use $- \kappa$
    \citep{Yoon-2014,Pierrad-etal-2016} as the exponent of the square bracket in Eq.~(\ref{eq:skd}) or $\kappa-\thalf$ 
    in the denominator \citep{Livadiotis-McComas-2013} in the square bracket in Eq.~(\ref{eq:skd}), with various 
    arguments relying in general on physical or theoretical implications of the velocity moments of Eq.~\ref{eq:skd}.
    
Despite its frequent successful employment, the SKD introduces certain unphysical characteristics. 
    The major deficiency of the SKD is the existence of diverging velocity moments, which prevents 
    establishing a fully consistent macroscopic non-equilibrium plasma model. For the $l$th velocity 
    moment to exist, $\kappa$ must fulfill the condition $\kappa > (l+1)/2$ \citep{Scherer-etal-2017d}. The definition of kinetic temperature from the second-order moment of the SKD restricts the spectral power to $\kappa > 3/2$. Furthermore, \cite{Scherer-etal-2019a} showed that for values of $\kappa < 2$ superluminal particles with $v > c$, (with $c$ the speed of light), contribute significantly to macroscopic quantities like the pressure or entropy. The concept of a non-additive entropy mentioned above is also still controversial 
    (see, e.g., the exchange between \cite{Nauenberg-2003, Tsallis-2004, Nauenberg-2004} and the discussion
    in \cite{Fichtner-etal-2018}). For the SKD there is also some discrepancy regarding the Debye length: 
    for example, while \cite{Mace-etal-1998} and \cite{Livadiotis-McComas-2014} derive a vanishing Debye length for $\kappa \to 3/2$, \cite{Treumann-etal-2004} find it to diverge in this limit, and \cite{Fahr-Heyl-2016} determine it to be ten times that of the associated Maxwellian plasma.
    
An important progress has been made by introducing the regularized $\kappa$-distribution (RKD),
    which admits a divergence-free macroscopic moment parameterization without any restriction for the 
    power-index $\kappa$ \citep{Scherer-etal-2017, Scherer-etal-2019a}. The RKD is defined as
\begin{align}\label{eq:rkd}
    f_{RKD}(\vec{r},\vec{v},t) =&
  \frac{n(\vec{r},t)}{\sqrt{\pi^{3}}\Theta^{3}}
  \frac{1}{\sqrt{\kappa^{3}} U\left(\frac{3}{2},\frac{3}{2}-\kappa,
  \alpha^{2}\kappa\right)}\\\nonumber
  & \hspace{1cm}\left(1 + \frac{v^{2}}{\kappa\Theta^{2}}\right)^{-(\kappa+1)}
   e^{-\alpha^{2} \frac{v^{2}}{\Theta^{2}}}\,,
\end{align}
    where $U(a,c,x)$ is the Kummer-$U$ or Tricomi function and $\alpha$ is the cutoff-parameter, 
    which reproduces the SKD for $\alpha = 0$. The characteristics and physical implications of 
    the RKD are still expolored, e.g., regarding the pressure and heat flux \citep{Lazar-etal-2019} 
    or the dispersion properties \citep{Husidic-etal-2020} in RKD-plasmas.

In literature we do find not only the standard/regularized $\kappa$-distributions, but also some 
    extensions. To date there is no consensus about a universally accepted version or interpretation of the 
    $\kappa$-distributions (see, e.g., \citet{Lazar-etal-2015} and \citet{Livadiotis-2015a} and references therein). In our present work 
    we unify all these attempts and propose a straightforward generalization of the
    $\kappa$-distributions, which we will call the $\kappa$-cookbook. From it all the already-known
    $\kappa$-distributions can be derived, which we will call recipes (see in Section~\ref{sec:cookbook} for 
    a detailed definition). Moreover, we can provide general analytical expressions for the velocity moments, 
    the Debye length and the entropy, and compare the results for the different recipes to analyze their 
    mathematical and physical significance. A further comparison can be achieved by fitting the recipes to real 
    data, for which we use an electron data set by the ESA space probe Ulysses from an event on February 15, 2002, 
    when Ulysses was in its second orbit around the Sun \citep{Marsden-Smith-2003}. We demonstrate that 
    the discussed recipes lead to almost the same values for the macroscopic moments and give easy to use 
    formulas for the velocity moments of the generalized kappa distributions, which can be used in future without 
    performing the explicit integration of the velocity moments.

The paper is organized as follows. In Section~\ref{sec:cookbook} we introduce a generalized 
    $\kappa$-distribution, which enables a rather general derivation of the moments, the Debye length, 
    and the entropy. Having these general formulations at hand, we discuss more properties in 
    Section~\ref{sec:prop}, while some specific recipes are analyzed in detail in 
    Section~\ref{sec:recipes}. In Section~\ref{sec:examples} we consider some examples 
    commonly used in the literature, and discuss the impact of choosing them as recipes of our 
    $\kappa$-cookbook. An examination of higher-order moments
    follows in Section~\ref{sec:moments}, while in Section~\ref{sec:DL} we take
    a closer look at the Debye length for each relevant recipe. Finally,
    we apply the discussed recipes to observations in Section~\ref{sec:appl},
    and end with a summary and conclusions in Section~\ref{sec:conclusion}.

\section{The \texorpdfstring{$\mathbf{\kappa}$}{kappa}-cookbook} \label{sec:cookbook}

Before we introduce the generalized $\kappa$-distribution (GKD), 
we briefly mention that all distribution functions
must obey the same physical laws, that are those derived from the
Liouville theorem \citep[e.g.,][]{Balescu-1988}. Hence the H-theorem
holds and the entropy is finite and extensive. Furthermore, we take it for granted that the
contribution of superluminal particles is negligible
\citep[see][]{Scherer-etal-2019b}. Moreover, it is required that the
Debye length $\Lambda$ is finite and positive. Thus, in short, a well-posed
distribution function in plasma physics must obey the following rules,
restricted to the case where the phase space volume is conserved:

\begin{itemize}
\item [1.] The distribution function should fulfill the Liouville
  theorem. Here we restrict ourselves to the case that it should be an (approximate)
  solution of the Vlasov equation \citep[e.g.\ ][]{Balescu-1988}.
\item [2.] All moments must exist \citep[e.g.\ ][]{Schwabl-2013}.
\item[3.] The entropy is given via the H-theorem
  \citep[e.g.][]{Balescu-1988}.
\item [4.] The plasma parameter $N_{p}=\frac{4\pi}{3}n_{0}\Lambda^{3}$
  must be high \citep[e.g.][]{Goedbloed-etal-2010}, where $N_p$ gives the 
  number of particles in a Debye sphere and $n_0$ is the number density.
\item [5.] The contribution of superluminal particles shall be
  negligible \citep{Scherer-etal-2019b}.
\end{itemize}
There can be other restrictions, but for our purposed dealing with
classical plasma distribution functions, the above conditions are necessary (see
also below).

In the following we study a generalized form of the isotropic regularized 
$\kappa$-distribution (RKD), which we  call the "cookbook": 
\begin{align}\label{eq:cook}
  f_{GKD}(\eta,\zeta,\xi,v) \equiv\, &n_{0}N_{G}
  \left(1+\frac{v^{2}}{\eta(\kappa)\Theta^{2}}\right)^{-\zeta(\kappa)}e^{-\xi(\kappa)\frac{v^{2}}{\Theta^{2}}}\\\nonumber
&  \qquad \mathrm{with\ }
   \eta(\kappa), \zeta(\kappa), \xi(\kappa)\in\mathbb{R}_{+}\,,
\end{align}
and specific values of
$\eta\equiv\eta(\kappa), \zeta\equiv\zeta(\kappa)$
and $\xi\equiv\xi(\kappa)$
are called the "recipes" given by the tuple $(\eta,\zeta,\xi)$. $N_{G}$ is the normalization constant (see below Eq.~\ref{eq:gnorm}).
We will discuss a selection of recipes below.

To save writing we introduce similar to our earlier definition \citep{Scherer-etal-2019a}
\begin{align}\label{eq:kummer_ratio}
  \Uka[m][n]\left(\eta,\zeta,\xi\right) =
  \frac{U\left(\frac{3+m}{2},\frac{5+m}{2}-\zeta,\xi\eta\right)}
  {U\left(\frac{3+n}{2},\frac{5+n}{2}-\zeta,\xi\eta\right)}\,.
\end{align}
where $n,m$ are arbitrary velocity moments. Because we only will need the ratios for $n=0$, we may further write
\begin{align}
  \Uka[m]\left(\eta,\zeta,\xi\right) =
  \frac{U\left(\frac{3+m}{2},\frac{5+m}{2}-\zeta,\xi\eta\right)}
  {U\left(\frac{3}{2},\frac{5}{2}-\zeta,\xi\eta\right)}\,.
\end{align}

First we give further ingredients of the $\kappa$-cookbook, 
namely the velocity moments, i.e., the $n$th order moment
$M_n$,
the normalization constant $N_{G}$, the most probable speed $v_{p}$,
and the pressure $P$ (see Appendix~\ref{integrals}). 

\begin{align}\label{eq:M_cook}\nonumber
    M_{n} &=  2\pi n_{0}N_{G}
          \eta^{\frac{3+n}{2}}\Theta^{3+n}\GA{\frac{n+3}{2}}
          U\left(\frac{n+3}{2},\frac{n+5}{2}-\zeta,\xi\eta\right)\\
  &=  \frac{2n_{0}}{\sqrt{\pi}}\eta^{\frac{n}{2}}\Theta^{n}\GA{\frac{n+3}{2}}
    \Uka[n](\eta,\zeta,\xi)
  \\\label{eq:gnorm}
  N_{G}^{-1} &= \frac{M_{0}}{n_{0}} = \eta^{}\Theta^{3}\sqrt{\pi^{3}} U\left(\thalf,\frac{5}{2}-\zeta,\xi\eta\right)\\\label{eq:gvp}
  v_{p}  = M_{1}  &= \frac{2 n_{0}\Theta}{\sqrt{\pi}}\eta^{\half}
\Uka[1](\eta,\zeta,\xi)
  \\\label{gpres}
  P = M_{2}  & = \frac{2 n_{0}\Theta^{2}}{3}\eta
\Uka[2](\eta,\zeta,\xi)\,.
\end{align}

With the approach by \citet{Krall-Trivelpiece-1973} (see
Appendix~\ref{subsec:krall_trivelpiece}) we estimate the \DL $\Lambda$ to
\begin{align}
      \Lambda^{2} & = \Lambda^{2}_{0} 
  \tilde{\Lambda}^{2} = \frac{1}{2} \Lambda^{2}_{0}\left[\frac{\zeta}{\eta}\frac{
    U\left(\thalf,\frac{3}{2}-\zeta,\eta\xi\right)}
    {U\left(\thalf,\frac{5}{2}-\zeta,\eta\xi\right)}
    + \xi  
    \right]^{-1}
\end{align}
with
\begin{align}\label{eq:lambda_0^2}
  \Lambda^{2}_{0} & =  \frac{\epsilon_0 m_{s}\Theta^{2}_{s}}{n_{s}
                    q_{s}^{2}}\,,\\\nonumber
  \tilde{\Lambda}^{2} &= \half \,\left[\frac{\zeta}{\eta}\frac{
    U\left(\thalf,\frac{3}{2}-\zeta,\eta\xi\right)}
    {U\left(\thalf,\frac{5}{2}-\zeta,\eta\xi\right)}
    + \xi  
    \right]^{-1}\,,
\end{align}
where the factor 2 comes from the fact that usually a Maxwellian is defined replacing $\Theta^{2}$  
with the thermal speed  and a factor 2: $\Theta^{2}=2 v_{p}^{2}$. 
In Eq.~\eqref{eq:lambda_0^2} $m_s$ denotes the mass and $q_s$ the charge of particle species 
$s$, and $\epsilon_0$ is the vacuum permittivity.

We will use the entropy $S$
(Appendix~\ref{entropy}) only in its normalized
version $\tilde{S}=S/(4\pi N k_{B})$
(with the Boltzmann constant $k_{B}$
and the total number of particles $N$).
By omitting the Gibbs correction
\citep[see for example][]{Fichtner-etal-2018}, we find for the entropy
\begin{align}\label{eq:entgkd2}
   \tilde{S} =& 
               -\ln n_{0} + 3 \ln \Theta +
                \ln\left[\eta^{\thalf}\sqrt{\pi^{3}}
      U\left(\thalf,\frac{5}{2}-\zeta,\xi\eta\right) \right]
               \\\nonumber
  &
  +\frac{\zeta}{2\pi} \sum\limits_{l=0}\frac{1}{l+1}\,
  \frac{U\left(\thalf,\thalf-\zeta-l,\eta\xi\right)}
  {U\left(\thalf,\frac{5}{2}-\zeta,\eta\xi\right)}
  +\frac{2\xi\eta}{3}\,
    \Uka[2](\eta,\zeta,\xi) \,.
\end{align}

From Eq.~(\ref{eq:entgkd2}) it is not directly evident that this 
non-equilibrium entropy is
lower than the classical equilibrium entropy discussed in
thermodynamics \citep[e.g.][]{Schwabl-2013}, but for the corresponding
Maxwellian we have to choose the correct temperature, e.g.,\ the one given
by the cookbook, and then it can be shown that it is lower than the
equilibrium entropy \citep[see][for the RKD case]{Scherer-etal-2019a}.

We can obtain the Maxwellian distribution function $f^{M}$ and the
corresponding moments $M_{n}^{M}$ with the recipe ($\eta>0,\zeta=0,\xi=1$) , for a
more detailed discussion see Section~\ref{sec:max}. It turns out that the general moments can be
written as a product of the Maxwellian moments and a ``correction''
factor $\tilde{M}$:
\begin{align}\label{gkdnorm}
  M_{n} = M_{n}^{M}\tilde{M} \qquad \mathrm{with}
\left\{\begin{array}{l}
  \tilde{M}\phantom{^{M}}\equiv \eta^{\frac{n}{2}}
\Uka[n]\\
M_{n}^{M} = \frac{2}{\sqrt{\pi}}\Theta^{n}\GA{\frac{n+3}{2}} \,.
\end{array}\right.
\end{align}
Analogously, we define $\tilde{v}_{p}$ and $\tilde{P}$ as the most
probable speed and pressure, respectively. In an analogous way we can
define the distribution function \begin{align}\label{maxwell-part}
  f = N_{M}\tilde{f} \qquad  \mathrm{with}
  \left\{\begin{array}{l}
    \tilde{f} \equiv
  \frac{1}{\eta^{\thalf}\sqrt{\pi^{3}}}
  \frac{\left(1+\frac{v^{2}}{\eta\Theta^{2}}\right)^{-\zeta}e^{-\xi\frac{v^{2}}{\Theta^{2}}}}
    {U\left(\thalf,\frac{5}{2}-\zeta,\eta\xi\right)}\\
    N_{M} = \frac{n_{0}}{\Theta^{3}}
     \end{array}\right.
\end{align}
with a "Maxwellian" normalization $N_M$.
In the following we do not take into account the Maxwellian part, but
rather study the normalized distribution functions and their moments.

\section{Some general properties}\label{sec:prop}
It is obvious that for $\{\eta(\kappa),\zeta(\kappa),\xi(\kappa)\}\in\mathbb{R}_{+}$ the
first three conditions are fulfilled, the fourth condition needs to be
checked, because it depends also on the number density, and the fifth
condition is fulfilled when $\xi>\frac{\Theta^{2}}{c^{2}}$, where
$c$ is the speed of light. More interesting
are the cases when one or more of the parameters ($\eta,\zeta,\xi$) are
zero. This can be the case if they are strictly zero, or if they
vanish for a given ($\kappa$-)value. First, we discuss the case when
they are strictly zero, because the other case can be deduced from the
former. In the last subsection, we study the recipes ($\eta \to
0,\zeta\ne 0, \xi\to \infty$), which can appear when replacing
$\xi$ by $\eta^{-1}\xi$ as used by \citet{DeStefano-2019}.  

In the following we discuss some special recipes, i.e.,\ $(\eta>0,\zeta=0,\xi\ne 0)$ 
the Maxwellian, $(\eta\ne 0,\zeta\ne 0,\xi=0)$ the SKD, and $(\eta=0,\zeta\ne 0,\xi\ne 0)$.
Furthermore, instead of using the shorthand notations SKD and RKD, we 
will mainly use the corresponding recipes. We study first some general properties of the 
GKD with finite and positive recipes, meaning that each parameter of tuple $(\eta,\zeta,
\xi)$ is finite and positive (Section~\ref{sec:form} below), and then the recipes, where 
one or more of the parameters vanishes (Section~\ref{sec:recipes}). 

\subsection{The form of the distribution function \texorpdfstring{$\tilde{f}$}{tilde f}}\label{sec:form}

\begin{figure*}\label{fig:shape}
  \includegraphics[width=.95\textwidth]{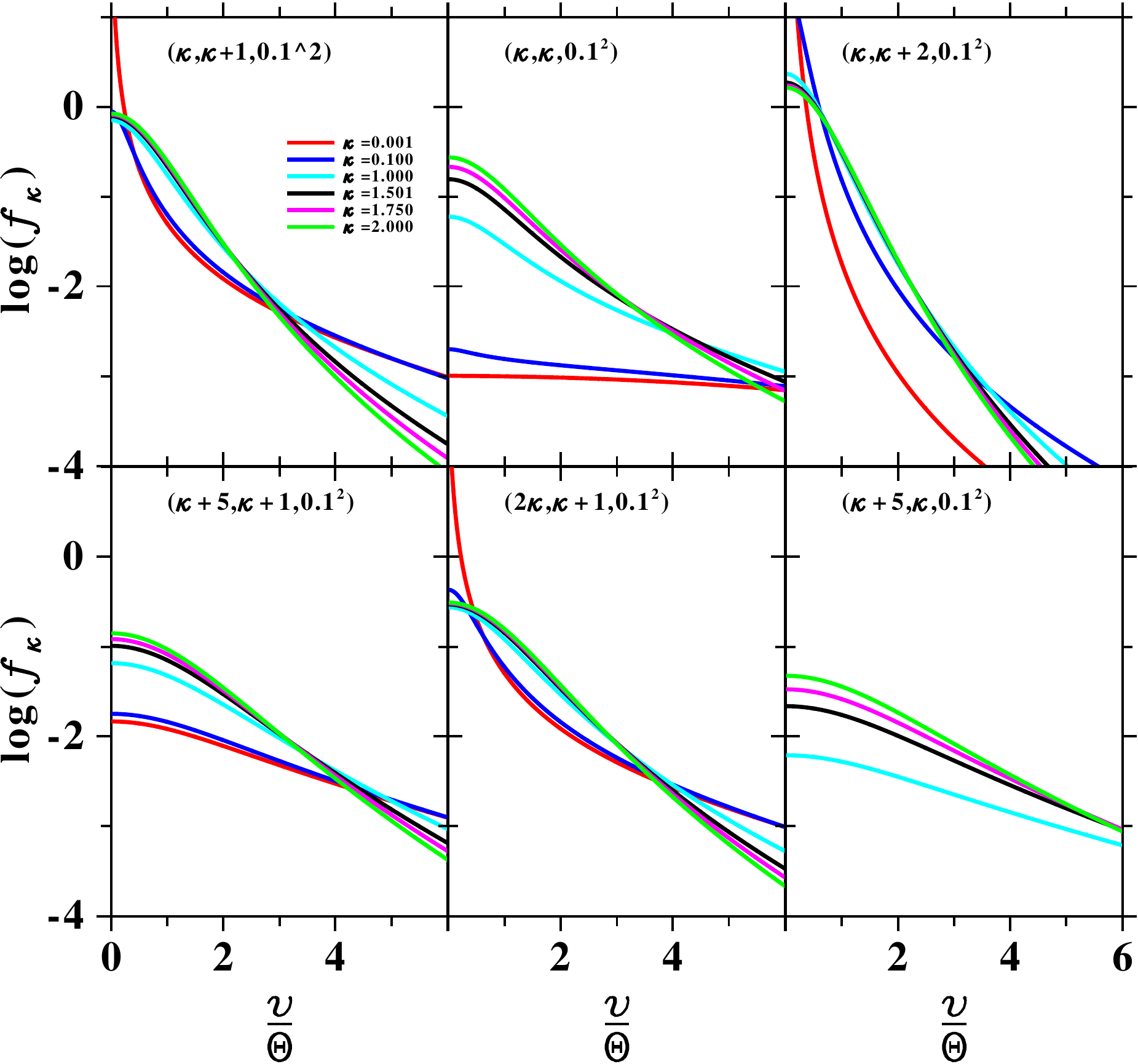}
  \caption{The shape and flatness parameters. The $\kappa$
    values can be found in the inlet of the left upper panel and are
    the same for all panels. For further discussion see main text.}
\end{figure*}

For this section we make the assumption that 
$\{\eta(\kappa),\zeta(\kappa),\xi(\kappa)\}\in\mathcal{R}_{+}$.
Inserting the normalization constant $N_{G}$ in Eq.~(\ref{eq:cook}) and
neglecting the Maxwellian part, see Eq.~(\ref{maxwell-part}),
leads to
\begin{align}
  \tilde{f} =
  \underbrace{\frac{1}{\eta^{\thalf}U(\thalf,\frac{5}{2}-\zeta,\eta\xi)}}_{{\mathrm{the
 \  normalization}}}
  \underbrace{\vphantom{\frac{1}{\eta^{\thalf}}}
  \left(1+\frac{v^{2}}{\eta(\kappa)\Theta^{2}}\right)^{-\zeta(\kappa)}}_{\mathrm{the
  \ form}}
  \underbrace{\vphantom{\frac{1}{\eta^{\thalf}}}
  e^{-\xi(\kappa)\frac{v^{2}}{\Theta^{2}}}}_{{\mathrm{the\
  cutoff}}}\,,
\end{align}
which has been decomposed into the above three parts. The form part
will be decomposed further in a part describing the tail flatness,
depending on $\zeta$, and a part describing the form of the peak. This
is illustrated in Fig.~\ref{fig:shape}, where we show in the left upper
panel the standard RKD, and in the upper middle and right panel we
vary $\zeta$ for $\eta=\kappa$, while in the lower left and  middle panel we vary $\eta$
for $\zeta=\kappa+1$, and finally in the lower right  panel we change
both $\zeta$ and $\eta$. The corresponding recipes are given in Table~\ref{tab:1}.
We have only plotted the low values of $v/\Theta$ so that the cutoff
parameter $\xi=0.1^{2}$ does not play a role, and thus the figures for
lower $\xi<0.1^{2}$ will look very similar and need not be included
in the above discussion.

Comparing the RKD 
with the recipe $(\kappa,\kappa,0.1^2)$
shows that the tails of the latter are much flatter than those of the
RKD and vice versa for the recipe $(\kappa,\kappa+2,0.1^{2})$,
which has steeper tails than the RKD. That is the reason why we call
the parameter $\zeta$ the flatness parameter.

Keeping $\zeta=\kappa+1$ and changing the $\eta$-parameter shows that
it influences the shape of the peak: The peaks for
the recipe $(\kappa+5,\kappa+1,0.1^{2})$
and for
$(2\kappa,\kappa+1,0.1^{2})$
are flatter than that of the RKD. Finally, in the
last panel we show how both parameters influence the form of the
recipe $(\kappa+5,\kappa,0.1^{2})$.

The discussed recipes (Table~\ref{tab:1}) or similar ones can be used
to fit a distribution function to data, but do not say anything about
the necessary physical conditions. This we will discuss in quite
general form in the next section. 

\section{Zero-recipes}\label{sec:recipes}
\subsection{The recipes with $\zeta=0$}\label{sec:max}

With the choice of $(\eta,\zeta,\xi)=(1,0,1)$
we get the standard Maxwellian distributions (up to a factor 2 when
comparing $\Theta$
with the thermal speed) and the corresponding moments from
Eq.~(\ref{eq:M_cook}):
 \begin{align}\label{eq:16}
     M_{n}^{M} &= \frac{2n_{0}}{\sqrt{\pi}} \Theta^{n}\GA{\frac{n+3}{2}}
         \underbrace{\frac{
          U\left(\frac{n+3}{2},\frac{n+5}{2},1\right)}
          { U\left(\frac{3}{2},\frac{5}{2},1\right)}}_{=1}
 \end{align}
(see Appendix, Eq.~(\ref{eq:aa})). Comparing this with the general moments we have always a part, which is
analogous to the Maxwellian moment, and we can write
\begin{align}
   M_{n} = M^{M}_{n}\eta^{\frac{n}{2}}\Uka[n](\eta,\zeta,\xi)\,.
\end{align}
Thus,  it is sufficient to discuss in the following only the
normalized moments
\begin{align}
\tilde{M}_{n}= \frac{M_{n}}{M^{M}}
\end{align}
and, analogously, for $\tilde{v}_{p}$ and $\tilde{P}$, and for the
distribution function $\tilde{f}$.

The recipes with $\eta\ne 1$ or $\xi\ne 1$ result in a scaling
of the Maxwellian and can easily be obtained from the above, though, 
they will not be discussed further.

\subsection{The $\xi=0$ recipes}\label{sec:zeta}

With the choice of $\xi=0$ the cutoff part vanishes and we are left
with 
\begin{align}
  \tilde{f}(\eta,\zeta,0) = \frac{1}{\eta^{\thalf}}
  \frac{\GA{\zeta}}{\GA{\zeta-\thalf}} 
  \left(1+\frac{v^{2}}{\eta\Theta^{2}}\right)^{-\zeta}\,,
\end{align}
where we have used Eq.~(\ref{eq:ulim}). The moments are
\begin{align}\label{eq:xi0mom}
  \tilde{M}_{n} = \eta^{\frac{n}{2}} \frac{\GA{\zeta-\frac{3+n}{2}}}{\GA{\zeta-\thalf}}
\qquad \qquad \mathrm{only\ if}\qquad \zeta > \frac{3+n}{2}\,.
\end{align}

If we check now for the contribution of particles beyond the speed of
light, i.e.,\ we calculate the relative pressure according to
\citet{Scherer-etal-2019b}, which is the ratio $R$ of the total thermal
pressure $\tilde{P}$ and the relative pressure $P'$:
\begin{align}
  R = 1 - \frac{P'}{P}
\end{align}
with
\begin{align}
  P' = \int\limits_{0}^{w} \tilde{f} v^{4}\md v\,,
\end{align}
where the integral is cut off at the upper boundary $w$. Thus, it is
easy to check that the ratio $R$ does not depend on $\eta$. Therefore,
we only need to estimate the integral in dependence of, 
say $\zeta_{0}$, and get the
contribution from superluminal particles for all distribution functions
$\tilde{f}(\eta,\zeta_{0},0)$. In the literature the function $\zeta$ is
determined by $\zeta=\kappa+1$, and hence the superluminal contribution
for $\eta=\kappa$ or $\eta=\kappa-\thalf$ is the same. We can also
state that with increasing flatness parameter
$\zeta(\kappa)>\zeta_{0}(\kappa),\ \forall \kappa$ the
contribution becomes lower than that for $\zeta_{0}=\kappa+1$ or higher
if $\zeta(\kappa)>\zeta_{0}(\kappa), \ \forall \kappa$.  

Thus, the recipes with $\xi=0$ have to be checked for the contribution
of superluminal particles, before they can be applied to physical
considerations. In the case that $\zeta=\kappa+1$, which is quite
often used in literature (see the references in the introduction), one
should not choose for sufficiently high values of $\Theta$ too low
values of $\kappa$, see the discussion in \citet{Scherer-etal-2019b}.

\subsection{$\eta=0$ recipes}
When this happens the distribution function 
from Eq.~(\ref{eq:cook}) goes to
infinity for $\kappa<\thalf$, or for $\kappa>\thalf$ to zero (see
Appendix~\ref{sec:eta0}) and all moments vanish.  
Moreover, we see from Eq.~(\ref{eq:xi0mom}) that the $n$th moment is
proportional to $\eta^{\frac{n}{2}}$ and therefore $\eta$ should be strictly
positive for non-vanishing moments. There is a
special case when we require for a given moment, say the $n$th one,
\begin{align}
  \eta^{\frac{n}{2}}
  \frac{\GA{\zeta-\frac{3+n}{2}}}{\GA{\zeta-\thalf}}
= C\,,
\end{align}
where $C$ is an arbitrary constant, which we can without loss of
generality choose as $C=1$, and if $C\ne1$ we divide $\eta$ by $C^{\frac{2}{n}}$. 
Thus we have
\begin{align}\label{zero}
  \eta^{\frac{n}{2}}\GA{\zeta-\frac{3+n}{2}}=\GA{\zeta-\thalf}\,.
\end{align}
This equation can only hold when $\GA{\zeta-\frac{3+n}{2}}$ goes to
(positive) infinity, because when $\eta$ goes to zero, the right hand side of
Eq.~(\ref{zero}) is always positive. The first value at which the
Gamma function on the left side goes to infinity is when
$\zeta=\frac{3+n}{2}$. Inserting that value in the Gamma function on
the right hand side leads to $\GA{\frac{n}{2}}$. If we find now a
solutions for $\eta_{1}$ at $n=n_{1}$, it is evident that we cannot find
other solution with $n\ne n_{1}$, because for another moment, say
$n_{2}$, Eq.~(\ref{zero}) has a different solution $\eta_{2}$, which is not possible,
because for all moments $\eta$ should be the same. Thus, we can find an
$n_{1}$ and $\eta_{1}$ for which Eq.~(\ref{zero}) is fulfilled when
$\eta_{1}\to 0$, but for all other $n$ the moments $\tilde{M}_{n\ne
    n_{1}}=0$ vanish.
That leads to a physically strange situation, where the moment
$\tilde{M}_{n=n_{1}}=1$ is one and all others $\tilde{M}_{n\ne
  n_{1}}=0$ are zero. For example if $n_{1}=2$, the second moment (i.e.,\ the
pressure) is one ($\tilde{M}_{2}=\tilde{P}=1$), but all other moments
vanish when $\eta_{1}\to 0$. That means we have a constant pressure
(temperature) but a vanishing most probable speed for $\eta_{1}\to
0$. Thus we have a gas with constant pressure but no internal motion,
which is physically not meaningful.  

The solution for $\eta^{\frac{n}{2}}$ can easily be determined 
\begin{align}\label{eq:zero1}
  \eta^{\frac{n}{2}} =
  \frac{\GA{\zeta-\thalf}}{\GA{\zeta-\frac{3+n}{2}}}\,.
\end{align}
Some results for $\zeta=\kappa$ and $\zeta=\kappa+1$ are given in Table~\ref{tab:zero}.
\begin{table}
  \centering
  \begin{tabular}{X|XX}
    n & \zeta & \eta \\
    \hline
    1 & \kappa  & \left(\frac{\GA{\kappa-\thalf}}{\GA{\kappa-2}}\right)^{2}\\
    1 & \kappa+1& \left(\frac{\GA{\kappa-\half}}{\GA{\kappa-1}}\right)^{2}\\
    2 & \kappa  &  \kappa-\frac{5}{2}\\
    2 & \kappa+1& \kappa - \frac{3}{2}\\
  \end{tabular}
  \caption{Solutions of Eq.~(\ref{eq:zero1}).}
  \label{tab:zero}
\end{table}

From Table~\ref{tab:zero} it becomes clear that the choice of
$\eta=\kappa-\thalf$ comes from the requirement that the second moment
for $\zeta=\kappa+1$ is constant and the temperature is equal to the
Maxwellian temperature ($M_{2}=P=M^{M}_{2}\tilde{M}=M^{M}_{2}$), which
has the deficit that all other moments vanish. 

Thus, to conclude the above discussion, the recipes with $\xi=0$ have a
couple of problems concerning their physical properties. The reason is
that when $\eta\to 0$, the distribution function
have a singularity  and all moments approach zero, except the one,
which is arbitrarily chosen to be constant.

\section{Some commonly used examples}\label{sec:examples}
We now discuss some recipes found in literature  (see
Table~\ref{choice}) or which were
suggested during our discussions. 
In the following we will always use $\zeta=\kappa+1$. The discussed
distribution functions  below can always be obtained by the proper choice of
$\xi(\kappa)$ and $\eta(\kappa)$ (see Table~\ref{choice}).

\begin{table}
  \begin{tabular}{l|XXX|l}
    name & \multicolumn{3}{c|}{recipe}&reference\\
    & \eta &\zeta&\xi &  \\
    \hline
    SKD & \kappa & \kappa+1 & 0 &\citep[e.g.,][]{Olbert-1968}\\
        &        &          &   &\citep{Vasyliunas-Siscoe-1976}\\
            &\kappa & \kappa+1 & 0& \citep[][not discussed here]{Yoon-2014}\\
         & \kappa-\thalf & \kappa+1 & 0&\citep[]{Livadiotis-McComas-2013}\\       & \kappa & \kappa+r & 0 & \citep{Treumann-Baumjohann-2014}\\    
    RKD & \kappa & \kappa+1 & \alpha^{2}& \citep[][]{Scherer-etal-2017}\\

        & \kappa-\thalf & \kappa+1 & \alpha^{2} & private communication\\
        & \kappa & \kappa+1 & \frac{\kappa\alpha^{2}}{\kappa-\thalf} \\
        & \kappa-\thalf & \kappa+1 & \frac{\kappa\alpha^{2}}{\kappa-\thalf}&\citep{DeStefano-2019}\\
  \end{tabular}
  \caption{The recipes $(\eta,\zeta,\xi)$
    for the discussed distribution functions, where $\alpha^{2}=\xi$
    is used to be compatible with our earlier notation.}
  \label{choice}
\end{table}

The complete definition of the distribution functions is given in
Table~\ref{tab:1}, where we have split the distribution functions in a
normalization part, which consists of a Maxwellian
(second column) and a remaining $\kappa$-part (third column), 
and a distribution part, which contains a power (fourth column) 
and a (possible) cutoff term (fifth column).
It can be seen from Table~\ref{tab:1} that the Maxwellian part
is the same for all distribution functions as already
discussed above.
The remaining normalization part in the third column
is a function of $\kappa$
and $\xi$ via the reciprocal of 
the Kummer-$U$ function, which is only the case if an
exponential part in the distribution exists (that is for all recipes
 $(\eta,\zeta,\xi)$ where $\xi\ne 0$). 

 The tail-part of the distribution given in the fourth column and the
 ``exponential'' cutoff in the fifth column describe the ``form'' of
 the distribution function.  The last two higher-ranking columns give
 the range, which is defined in such a way that the distribution
 function is always in $\mathbb{R}_{+}$.
 For the recipe $(\kappa,\kappa+1,0)$
 (SKD) it is in principle possible to have lower $\kappa$
 values than $\thalf$,
 but then the pressure is not defined (see below). Therefore, we
 choose for the SKD also as the lower limit $\kappa=\thalf$.
 Nevertheless, if we allow lower $\kappa$
 values in the SKD, we find that the lower limit is $\kappa=\half$
 at which the distribution vanishes. Thus, lower limits for the SKD
 and for the recipes $(\kappa-\thalf,\kappa+1,0)$,
 $(\kappa-\thalf,\kappa+1,\xi^{2})$,
 $(\kappa,\kappa+1,\frac{\kappa\xi^{2}}{\kappa-\thalf})$,
 are always at $\kappa=\thalf$,
 and that for the RKD at $\kappa=0$.
 At that limit the SKD has finite values depending on the
 velocity. All other distribution functions (including the RKD) go to
 infinity, when $v=0$
 (column 6), while for $v\ne 0$
 (column 7) only the RKD is a function of the velocity, while all the
 other discussed recipes tend to zero.  The eighth column gives the upper
 limits, when $\kappa \to \infty$.
 In this limit the the recipes with $\xi=\const$
 or $\xi=0$
 approach a Maxwellian type distribution, while the recipes with
 $\xi\propto (\kappa-\thalf)^{-1}$
 tend to zero. The upper limits are calculated using the integral of
 the 0th order moment to determine the normalization factors when
 $\kappa\to\infty$.
 The above recipes are described in Table~\ref{choice} with the
 corresponding reference. For further use we will call all the
 recipes with $\kappa-\thalf$
 dependency in one or more parameter the $\Psi$-distributions.

The range of the SKD could be extended to
$\kappa \in (\half\ldots\infty)$,
but then neither the most probable speed nor the pressure are
defined. All of the above distributions go to infinity when first
$v\to0$
and then $\kappa\to\thalf$, or $\kappa\to0$ in the case of the
RKD. Therefore, we calculated two limits for $\kappa\to
0$:  one for $v=0$ and one for $v\ne 0$. The calculations of the
limits for the Kummer-$U$ function are presented in Appendix~\ref{limits}.
  
\begin{table*}
 \scalebox{0.9}{ \begin{tabular}{L|L@{}K|LL|KK|K}
       \multicolumn{1}{c|}{recipe}
    & \multicolumn{2}{c|}{Normalization}  
    & \multicolumn{2}{c|}{Distribution}  
    & \multicolumn{2}{c|}{lower Limit}  
    & \multicolumn{1}{c}{upper  Limit}  \\
       \multicolumn{1}{c|}{}
    & \multicolumn{1}{c}{Maxwellian}
    & \multicolumn{1}{c|}{$\kappa$}
    & \multicolumn{1}{c}{tail}
    & \multicolumn{1}{c|}{cutoff}
    & \multicolumn{1}{c}{$\lim\limits_{\substack{\kappa\to 0 \\v=0}} =$}
    & \multicolumn{1}{c|}{$\lim\limits_{\substack{\kappa\to 0 \\v\ne 0}} =$}
    & \multicolumn{1}{c}{$\lim\limits_{\kappa\to \infty} =$}\\
   \hline
(\eta,\zeta,\xi^{2})
    &  \frac{n_{0}}{\sqrt{\pi^{3}}\Theta^{3}}
    & \frac{1}{ \eta^{\thalf}(\kappa)}\Uka[][0]\left(\kappa,\zeta,\eta\xi\right)
    & \left(1 + \frac{v^{2}}{\eta(\kappa)\Theta^{2}}\right)^{-\zeta(\kappa)}
    &  e^{-\xi(\kappa)\frac{v^{2}}{\Theta^{2}}}
    & -
    & -
    & -\\
    \hline
(\kappa, \kappa+1,0)
  &  \frac{n_{0}}{\sqrt{\pi^{3}}\Theta^{3}}
  & \frac{\Gamma(\kappa)}{\sqrt{\kappa}\Gamma\left(\kappa-\frac{1}{2}\right)}
  &\left(1 + \frac{v^{2}}{\kappa\Theta^{2}}\right)^{-\kappa-1}
  &
  & \infty\hfill~
  & \frac{n_{0}}{\sqrt{6}\pi}
    \left(1 + \frac{2}{3}
    \frac{v^{2}}{\Theta^{2}}\right)^{-\frac{5}{2}}
  &  \frac{n_{0}}{\sqrt{\pi^{3}}\Theta^{3}}
    e^{-\frac{v^{2}}{\Theta^{2}}}
    \hfill~
    \\
 (\kappa, \kappa+1, \xi^{2})
  & \frac{n_{0}}{\sqrt{\pi^{3}}\Theta^{3}}
  & \frac{1}{\kappa^{\thalf}}\
    \Uka[][0](\kappa,\kappa+1,\xi^{2})
  & \left(1 + \frac{v^{2}}{\kappa\Theta^{2}}\right)^{-\kappa-1}
  &  e^{- \frac{\xi^{2}v^{2}}{\Theta^{2}}}
  & \infty\hfill~
  &  \half \frac{n_{0}\xi } {\sqrt{\pi^{3}}\Theta v^{2}} e^{-\xi^{2}  \frac{v^{2}}{\Theta^{2}}}
  &  \frac{n_{0}\left(\xi^{2}+1\right)^{\frac{3}{2}}}{\sqrt{\pi^{3}}\Theta^{3}}
    e^{-(1+\xi^{2}) \frac{v^{2}}{\Theta^{2}}}\hfill~
  \\
  \hline
 (\kappa-\thalf, \kappa+1,0)
    &  \frac{n_{0}}{\sqrt{\pi^{3}}\Theta^{3}}
    &\frac{\kappa\Gamma(\kappa)}{\Gamma\left(\kappa-\frac{1}{2}\right)\left(\kappa-\frac{3}{2}
      \right)^{\frac{3}{2}}} 
    &\left(1 +
      \frac{v^{2}}{\left(\kappa-\frac{3}{2}\right)\Theta^{2}}\right)^{-\kappa-1}
    & \
    & \infty\hfill~
    & \quad\quad\quad 0 \hfill~
    & \frac{n_{0}}{\sqrt{\pi^{3}}\Theta^{3}} e^{-\frac{v^{2}}{\Theta^{2}}} \hfill~
  \\
 (\kappa-\thalf, \kappa+1, \xi^{2})
  & \frac{n_{0}}{\sqrt{\pi^{3}}\Theta^{3}}
  & \frac{1}{\left(\kappa-\frac{3}{2}\right)^{\frac{3}{2}}} 
    \Uka[][0]\left(\kappa-\thalf,\kappa+1,\xi^{2}\right)       
  & \left(1 + \frac{v^{2}}{\left(\kappa-\frac{3}{2}\right)\Theta^{2}}\right)^{-\kappa-1} 
  &  e^{-\xi^{2} \frac{v^{2}}{\Theta^{2}}}
  & \infty\hfill~
  & \quad\quad\quad 0 \hfill~
  & \frac{n_{0}\left(\xi^{2}+1\right)^{\frac{3}{2}}}{\sqrt{\pi^{3}}\Theta^{3}}
    e^{-(1+\xi^{2}) \frac{v^{2}}{\Theta^{2}}}
  \\
 (\kappa, \kappa+1, \frac{\kappa\xi^{2}}{\kappa-\thalf})
  & \frac{n_{0}}{\sqrt{\pi^{3}}\Theta^{3}}
  & \frac{1}{\kappa^{\frac{3}{2}}}
    \Uka[][0]\left(\kappa,\kappa+1,\frac{\kappa\xi^{2}}{\kappa-\thalf}\right)
  & \left(1 + \frac{v^{2}}{\kappa\Theta^{2}}\right)^{-\kappa-1}
  &  e^{-\frac{\kappa\xi^{2} v^{2}}{\left(\kappa-\frac{3}{2}\right)\Theta^{2}}} 
  &  \infty\hfill~
  & \quad\quad\quad 0 \hfill~
  &  \frac{n_{0}\left(\xi^{2}+1\right)^{\frac{3}{2}}}{\sqrt{\pi^{3}}\Theta^{3}}
    e^{-(1+\xi^{2}) \frac{v^{2}}{\Theta^{2}}}
  \\
 (\kappa-\thalf, \kappa+1, \frac{\kappa\xi^{2}}{\kappa-\thalf})
  & \frac{n_{0}}{\sqrt{\pi^{3}}\Theta^{3}}
  & \frac{1}{\left(\kappa-\frac{3}{2}\right)^{\frac{3}{2}}}
    \Uka[][0]\left(\kappa,\kappa+1,\frac{\kappa\xi^{2}}{\kappa-\thalf}\right)
  & \left(1 + \frac{v^{2}}{\left(\kappa-\frac{3}{2}\right)\Theta^{2}}\right)^{-\kappa-1}
  &  e^{- \frac{\kappa\xi^{2}v^{2}}{\left(\kappa-\frac{3}{2}\right)\Theta^{2}}} 
  & \infty\hfill~
  & \quad\quad\quad 0\hfill~
  &  \frac{n_{0}\left(\xi^{2}+1\right)^{\frac{3}{2}}}{\sqrt{\pi^{3}}\Theta^{3}}
    e^{-(1+\xi^{2}) \frac{v^{2}}{\Theta^{2}}}
 \end{tabular}
}
\caption{The regularized and quasi-regularized distribution
  function\label{tab:1}. The range for all distributions is assumed to
  be $\kappa\in \left(\left.\thalf \ldots\infty\right)\right.$,
  except for the RKD, in which case it goes from
  $\kappa\in (0 \ldots\infty)$.  }
\end{table*}

\begin{figure*}
  \centering
  \includegraphics[width=1.0\textwidth]{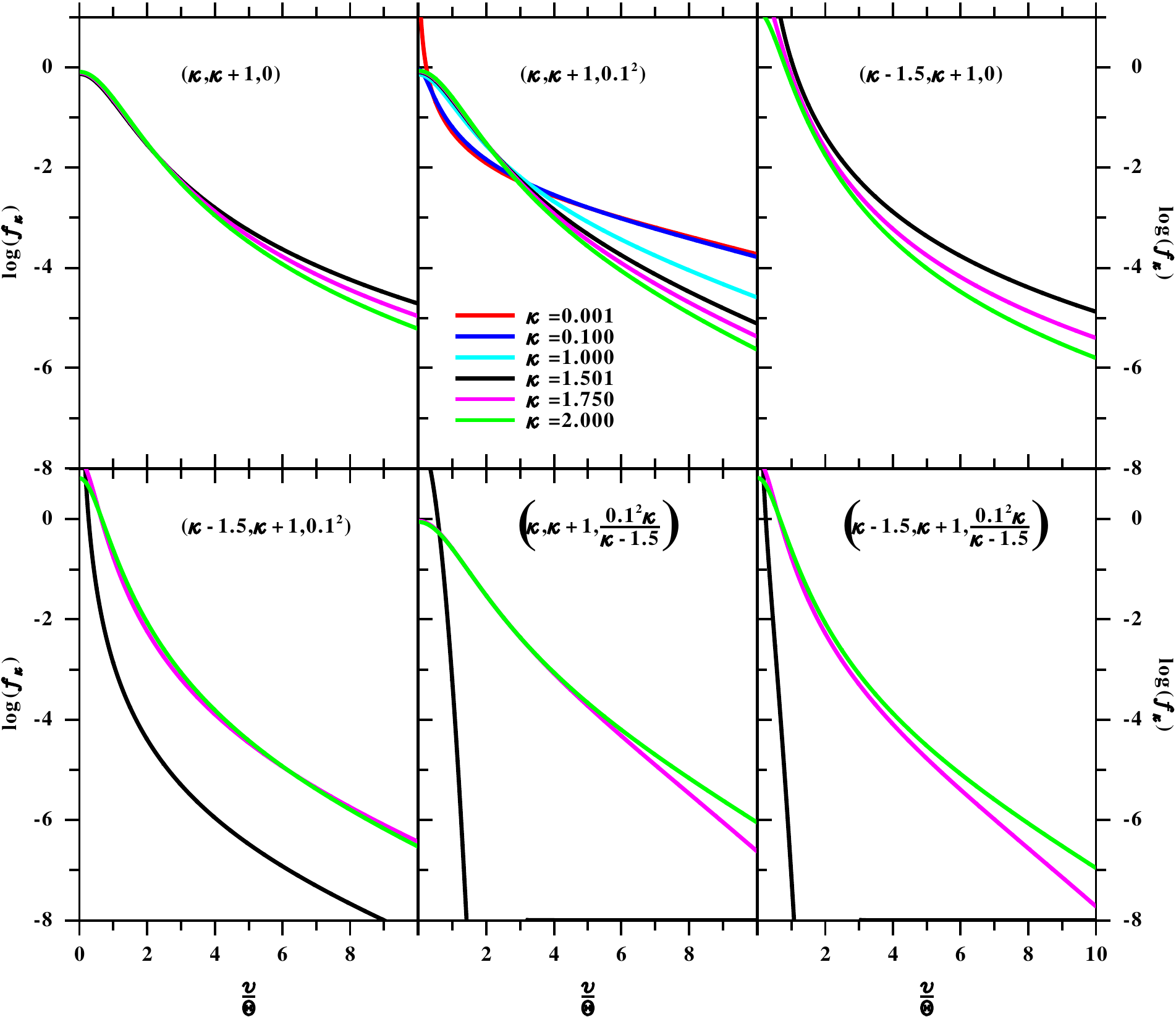}
  \caption{The different normalized distribution functions $f_{p}$
    for $\xi=10^{-1}$.
    The color codes are the same for all six
    panels.  
    Only for the recipe $(\kappa,\kappa+1,\xi^{2})$
    (RKD) the first three $\kappa$-values
    are meaningful. It can be seen that only the recipes
    $(\kappa, \kappa+1,0)$
    (SKD) and
    $(\kappa, \kappa+1, \frac{\kappa\xi^{2}}{\kappa-\thalf})$
    have finite values when $\kappa\to\thalf$,
    while the recipes $(\kappa-\thalf, \kappa+1, 0)$,
     $(\kappa-\thalf, \kappa+1, \xi^{2})$,
    and
    $(\kappa-\thalf, \kappa+1, \frac{\kappa\xi^{2}}{\kappa-\thalf})$
    go to infinity when $\kappa\to\thalf$.
    The RKD is also defined for all values $\kappa>0$.
    It can also be seen that the recipes
    $(\kappa-\thalf, \kappa+1, 0)$,
     $(\kappa-\thalf, \kappa+1, \xi^{2})$,
     $(\kappa, \kappa+1, \frac{\kappa\xi^{2}}{\kappa-\thalf})$, and
    $(\kappa-\thalf, \kappa+1, \frac{\kappa\xi^{2}}{\kappa-\thalf})$
    fall faster to zero with increasing $\kappa$ than the SKD and RKD.
  \label{fig:0a}}
\end{figure*}

The limits of the Kummer-$U$ function are discussed in Appendix~\ref{limits}. 
The calculation of the normalization factors given in columns 2 and 3 of
Table~\ref{tab:1} are given in the Appendix~\ref{integrals}. The \KUT
$U\left(\frac{n+3}{2},\frac{n+3}{2}-\kappa,x\right)$
is for arbitrary $x$
approximated as $U\left(\frac{n+3}{2},\frac{n+3}{2},x\right)$
if $\kappa\to 0$,
and as $U\left(\frac{n+3}{2},\frac{n}{2},x\right)$
if $\kappa\to \thalf$. These approximations are quite good for
$\kappa\ll 1$, as can be
seen in Fig.~\ref{fig:0a}, where the non-approximated functions are
plotted.

In Fig.~\ref{fig:0a} the distribution functions corresponding
to six recipes are plotted for
$\xi=0.1$.
For the RKD for $\kappa$-values 
of $\kappa \in \{0.001, 0.1, 1.0, 1.501, 1.75, 2.0 \}$
while for the other five distributions only the values
$\kappa \in \{1.501, 1.75, 2.0 \}$
are presented. In the upper panels the SKD, RKD and recipe $(\kappa-\thalf, \kappa+1, 0)$ are shown,
while in the lower panels the recipes $(\kappa-\thalf, \kappa+1, \xi^{2})$,  $(\kappa, \kappa+1, \frac{\kappa\xi^{2}}{\kappa-\thalf})$, and $(\kappa-\thalf, \kappa+1, \frac{\kappa\xi^{2}}{\kappa-\thalf})$ are presented. It
can be seen that the $\Psi$-
distributions decrease much faster with increasing $\frac{v}{\Theta}$
than the SKD and the RKD. For the limiting values ($\kappa=\thalf$
for the SKD and $\Psi$
distributions, and $\kappa=0$
for the RKD) all distribution functions tend to infinity, when $v\to 0$,
except the SKD, which has a finite value for $v=0$.
For sufficiently high $\kappa$
values we find a regular behavior for all distribution functions. Also,
the strange behavior for the recipes $(\kappa, \kappa+1, \frac{\kappa\xi^{2}}{\kappa-\thalf})$ and  $(\kappa-\thalf, \kappa+1, \frac{\kappa\xi^{2}}{\kappa-\thalf})$ can be seen
when $\kappa=1.501$:
For $v=0$
these distributions tend to infinity, but for $v\ne 0$
they rapidly decrease towards zero. This is the same behavior as shown
in Table~\ref{tab:1}. Also the recipe $(\kappa-\thalf, \kappa+1, 0)$ and recipe $(\kappa-\thalf, \kappa+1, \xi^{2})$ tend to zero for $v\ne 0$,
but this decrease is much weaker for $\kappa=1.501$.
The reason for this different behavior is the exponential term in the
recipes $(\kappa, \kappa+1, \frac{\kappa\xi^{2}}{\kappa-\thalf})$ and  $(\kappa-\thalf, \kappa+1, \frac{\kappa\xi^{2}}{\kappa-\thalf})$, which goes quickly to zero, when in the exponential term
$\kappa\to\thalf$.

\begin{figure*}
  \centering
  \includegraphics[width=1.0\textwidth]{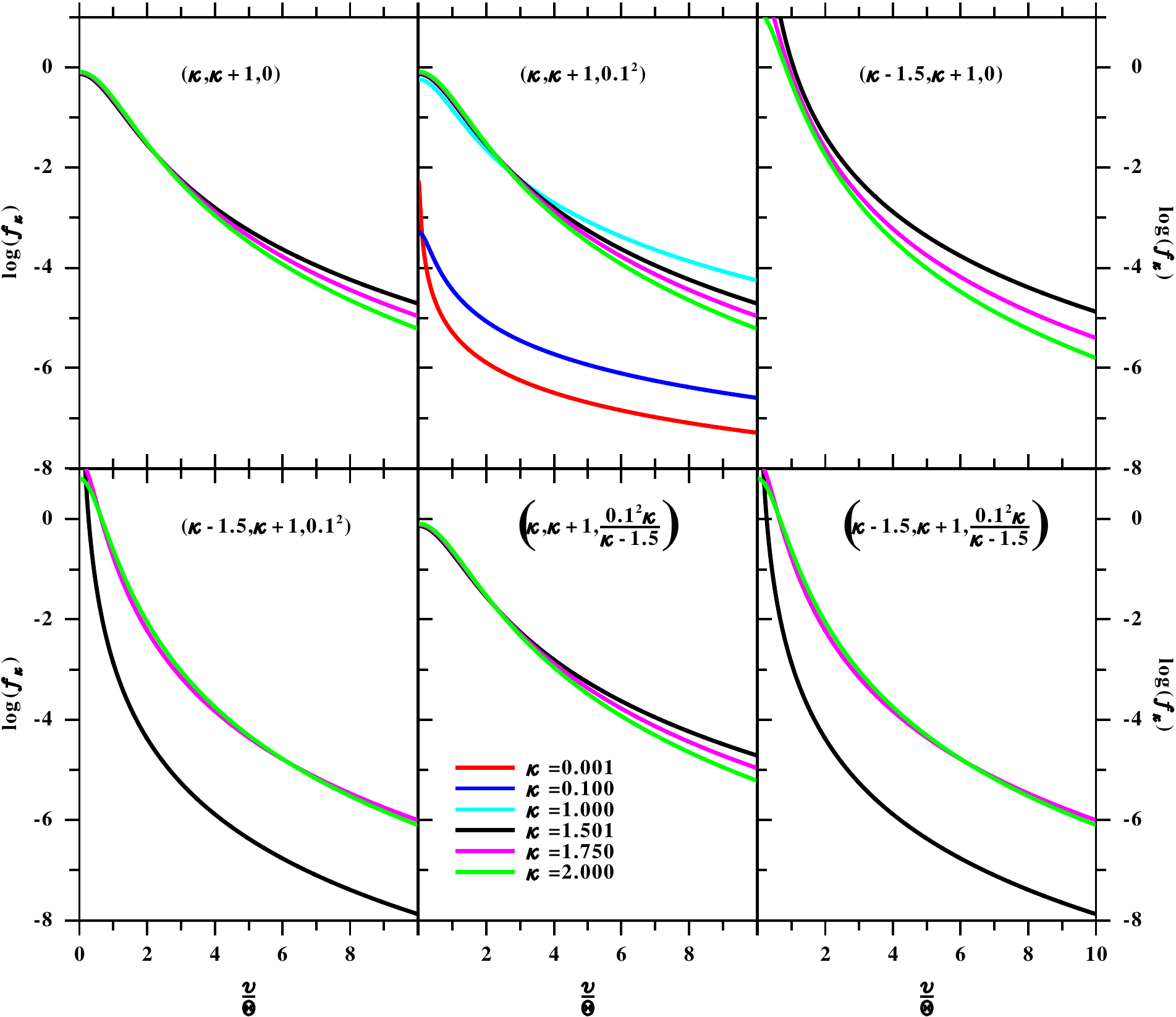}
  \caption{The different normalized distribution functions
    $f_{p}$ for $\xi=10^{-5}$. The colors are the same as in
    Fig.~\ref{fig:0a}. It can be seen that the RKD has flatter tails 
    for values $\kappa<1$, but its absolute value is much below
    the $\kappa>1$-values.
  \label{fig:0b}}
\end{figure*}

A few words to the RKD: For $\kappa$-values below $\kappa=\thalf$ the
tails become flatter up to high ratios of
$\frac{v}{\Theta}$. The tails for $\kappa=0.001$ and
$\kappa=0.1$ are almost equal, while the latter has reasonable values
at $v\approx0$.

In Fig.~\ref{fig:0b} we changed the regularization parameter to 
$\xi=10^{-5}$,
which does not affect the SKD and recipe $(\kappa-\thalf, \kappa+1, 0)$, but the other
distributions. The recipe $(\kappa-\thalf, \kappa+1, \xi^{2})$ and recipe $(\kappa-\thalf, \kappa+1, \frac{\kappa\xi^{2}}{\kappa-\thalf})$ are more relaxed for low
$\kappa$-values, because the low $\xi$ cancels the steep cutoff
at low values of $\frac{v}{\Theta}$, but will steeply fall off for high
enough $\kappa$-values. The recipe $(\kappa, \kappa+1, \frac{\kappa\xi^{2}}{\kappa-\thalf})$ behaves similar to the SKD, but with a
cutoff (not shown). The RKD has now for low $\kappa$-values lower
finite values at $v\approx 0$ and again flat tails for low
$\kappa$-values. Also, for such low $\xi$ values the recipes $(\kappa-\thalf, \kappa+1, \xi^{2})$ and  $(\kappa-\thalf, \kappa+1, \frac{\kappa\xi^{2}}{\kappa-\thalf})$
fall faster towards zero than the SKD and the RKD (the recipe $(\kappa-\thalf, \kappa+1, 0)$ does also,
but it depends not on $\xi$).

One can also see, especially from Fig.~\ref{fig:0a} that for the RKD
the Maxwellian core becomes lower, when the tails become flatter,
i.e.,\ the $\kappa$-values are low. This can be understood, because
the number of particles is constant and thus, if there are more
particles in the tail, less particles can be in the Maxwellian core.

\section{Higher-order moments}\label{sec:moments}

In Table~\ref{tab:3} the higher-order moments, i.e.,\ the most probable speed
(1. moment) and the pressure (2. moment) are presented together with
the \DL. The corresponding  integrals can be found in the
Appendices~\ref{integrals} and~\ref{Debye}. The \DL will be
discussed separately in section~\ref{sec:DL}.  

Again, we find that the most probable speed and pressure can be
decomposed into a Maxwellian part and a $\kappa$-dependent part. 
Thus, we will use the short hand notation:
$\tilde{u}_{p}= u_{p}\frac{\sqrt{\pi}}{2n_{0}\Theta}$
and $\tilde{P}= P\frac{3}{2n_{0}\Theta^{2}}$. This holds also true for
the \DL, which has a Maxwellian-like and a $\kappa$-part:
$\tilde{\Lambda}^{2} =
\Lambda^{2}\frac{q^{2}n_{0}}{\epsilon_{0}m\Theta^{2}}=
\frac{\Lambda^{2}}{\Lambda_{0}^{2}}$, where $\Lambda_{0}^{2}=
\frac{\epsilon_{0}m\Theta^{2}}{q^{2}n_{0}}$ is similar to the
Maxwellian \DL, except that the core speed $\Theta$ is used instead of
the Maxwellian thermal speed $v_{th}$. Here, $q$ is the elementary
charge, $\epsilon_{0}$ the electric permittivity for vacuum, and $m$ the
particle mass in mind.

 \begin{table*}[t!]
   \scalebox{0.90}{
  \begin{tabular}{l|L|K|K@{}K|K|K@{}K|X}	
    \multicolumn{1}{c|}{name}
    &\multicolumn{1}{c|}{recipe}
    &\multicolumn{3}{c|}{most probable Speed $u_{p}$}
    &\multicolumn{3}{c|}{Pressure $P$}
    &\multicolumn{1}{c}{\DL $\tilde{\Lambda}^{2}$}
    \\
  \hline  
GKD&(\eta,\zeta,\xi) & \frac{2 n_{0}}{\sqrt{\pi}} \Theta
    & \eta^{\half}(\kappa)
    & \Uka[1][0](\eta,\zeta,\xi)
    &  \frac{2 n_{0}}{3} \Theta^{2}
    & \eta(\kappa)
    & \Uka[2][0](\eta,\zeta,\xi)
& \left(\frac{\zeta}{\eta}
  \frac{U\left(\thalf,\thalf-\zeta,\eta\xi\right)}
  {U\left(\thalf,\frac{5}{2}-\zeta,\eta\xi\right)} +
  \xi 
  \right)^{-1}
    \\
    \hline
 SKD& (\kappa,\kappa+1,0) &   \frac{2 n_{0}}{\sqrt{\pi}} \Theta 
      &  \sqrt{\kappa}
      &   \frac{\GA{\kappa-1}}{\GA{\kappa-\frac{1}{2}}}
      & \frac{2 n_{0}}{3} \Theta^{2}
      & \kappa
      &\frac{1}{\kappa-\thalf}
      & \frac{\kappa}{\kappa-\half}
   \\
 RKD& (\kappa,\kappa+1,\xi^{2}) & \frac{2 n_{0}}{\sqrt{\pi}} \Theta
      & \sqrt{\kappa}
      &  \Uka[1][0](\kappa,\kappa+1,\xi^{2})
      & \frac{2 n_{0}}{3} \Theta^{2}
      &  \kappa
      & \Uka[2][0](\kappa,\kappa+1,\xi^{2})
    & \left[\frac{\kappa+1}{\kappa}
      \frac{U\left(\thalf,\half-\kappa,\kappa\xi^{2}\right)}
      {U\left(\thalf,\thalf-\kappa,\kappa\xi^{2}\right)}
      + \xi^{2}
      \right]^{-1}
 \\
  \hline
  &(\kappa-\thalf, \kappa+1, 0) &  \frac{2 n_{0}}{\sqrt{\pi}} \Theta
      & \left(\kappa -\frac{3}{2}\right)^{\frac{1}{2}}
      & \frac{\GA{\kappa-1}}{\GA{\kappa-\frac{1}{2}}}
      & \frac{2 n_{0}}{3} \Theta^{2}  
      & 1
      & 1
      & \frac{\kappa-\thalf}{\kappa-\half}
 \\
&(\kappa-\thalf, \kappa+1, \xi^{2}) & \frac{2 n_{0}}{\sqrt{\pi}} \Theta                                        
      & \left(\kappa -\frac{3}{2}\right)^{\frac{1}{2}}
      & \Uka[1][0]\left(\kappa-\thalf,\kappa+1,\xi^{2}\right)
      &  \frac{2 n_{0}}{3} \Theta^{2}  
      & \left(\kappa -\frac{3}{2}\right)
      &  \Uka[2][0]\left(\kappa-\thalf,\kappa+1,\xi^{2}\right)
      &\left[\frac{\kappa+1}{\kappa-\thalf}
      \frac{U\left(\thalf,\half-\kappa,(\kappa-\thalf)\xi^{2}\right)}
      {U\left(\thalf,\thalf-\kappa,(\kappa-\thalf)\xi^{2}\right)}
      + \xi^{2}
      \right]^{-1}
 \\
 &(\kappa, \kappa+1, \frac{\kappa\xi^{2}}{\kappa-\thalf}) &  \frac{2 n_{0}}{\sqrt{\pi}} \Theta
      &  \sqrt{\kappa}
      & \Uka[1][0]\left(\kappa,\kappa+1,\frac{\kappa\xi^{2}}{\kappa-\thalf}\right)
      &  \frac{2 n_{0}}{3} \Theta^{2}  
      & \kappa
      & \Uka[2][0]\left(\kappa,\kappa+1,\frac{\kappa\xi}{\kappa-\thalf}\right)
      & \left[\frac{\kappa+1}{\kappa}
      \frac{U\left(\thalf,\half-\kappa,\frac{\kappa^{2}\xi^{2}}{\kappa-\thalf}\right)}
      {U\left(\thalf,\thalf-\kappa,\frac{\kappa^{2}\xi^{2}}{\kappa-\thalf}\right)}
      + \frac{\kappa\xi^{2}}{\kappa-\thalf}
      \right]^{-1}
    \\
 &(\kappa-\thalf, \kappa+1, \frac{\kappa\xi^{2}}{\kappa-\thalf}) &  \frac{2 n_{0}}{\sqrt{\pi}} \Theta
      & \left(\kappa -\frac{3}{2}\right)^{\frac{1}{2}}
      & \Uka[1][0]\left(\kappa-\thalf,\kappa+1,\frac{\kappa\xi^{2}}{\kappa-\thalf}\right)
      &  \frac{2 n_{0}}{3} \Theta^{2}  
      & \left(\kappa -\frac{3}{2}\right)
      &\Uka[2][0]\left(\kappa-\thalf,\kappa+1,\frac{\kappa\xi^{2}}{\kappa-\thalf}\right)
      & \left[\frac{\kappa+1}{\kappa-\thalf}
      \frac{U\left(\thalf,\half-\kappa,\kappa\xi^{2}\right)}
      {U\left(\thalf,\thalf-\kappa,\kappa\xi^{2}\right)}
      + \frac{\kappa\xi^{2}}{\kappa-\thalf}
      \right]^{-1}
\end{tabular}
}
\caption
{The most probable speeds, pressures and the normalized Debye lenghts \label{tab:3}.} 
\end{table*}

\begin{figure*}
  \centering
  \includegraphics[width=0.45\textwidth]{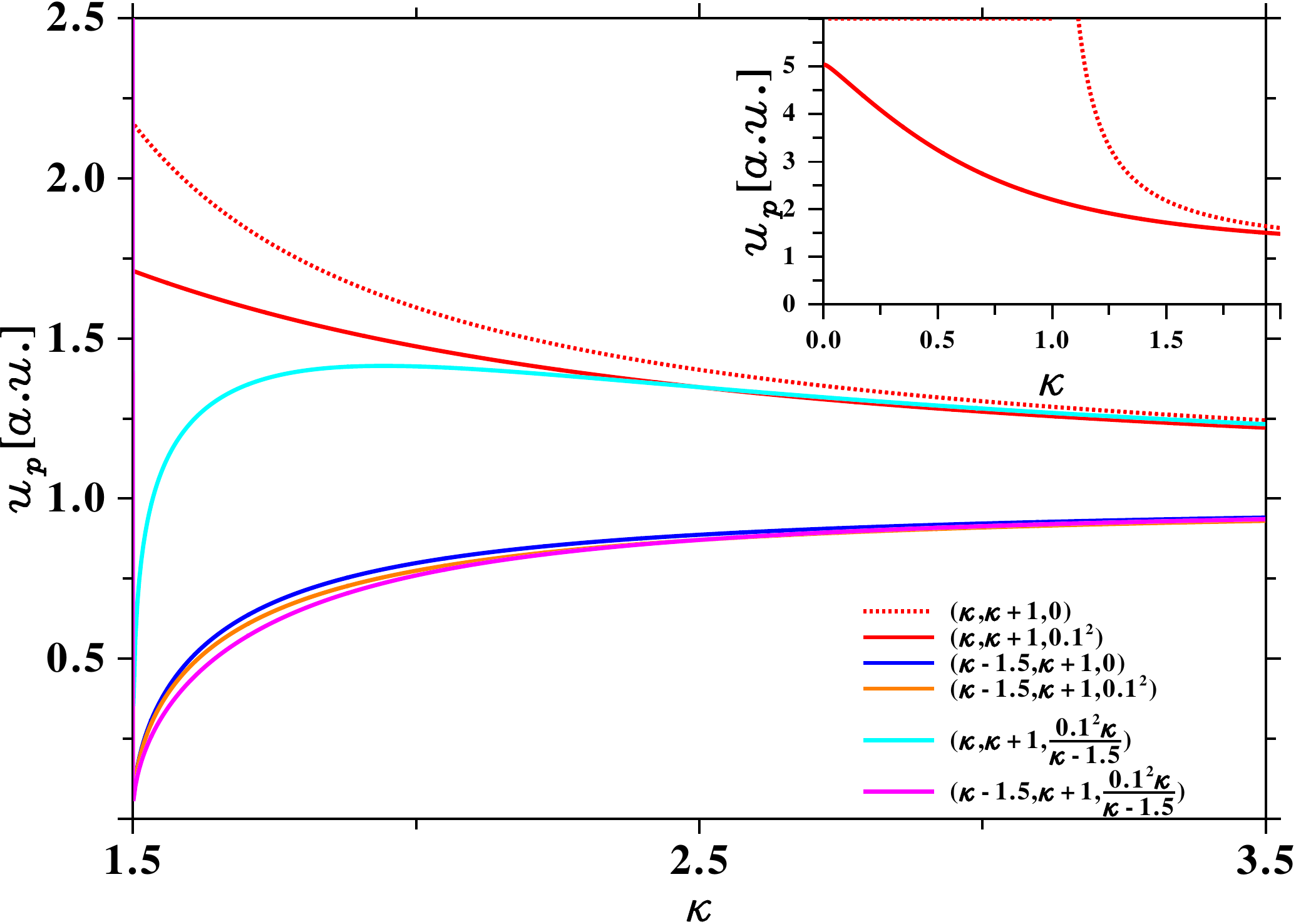}\hfill
  \includegraphics[width=0.45\textwidth]{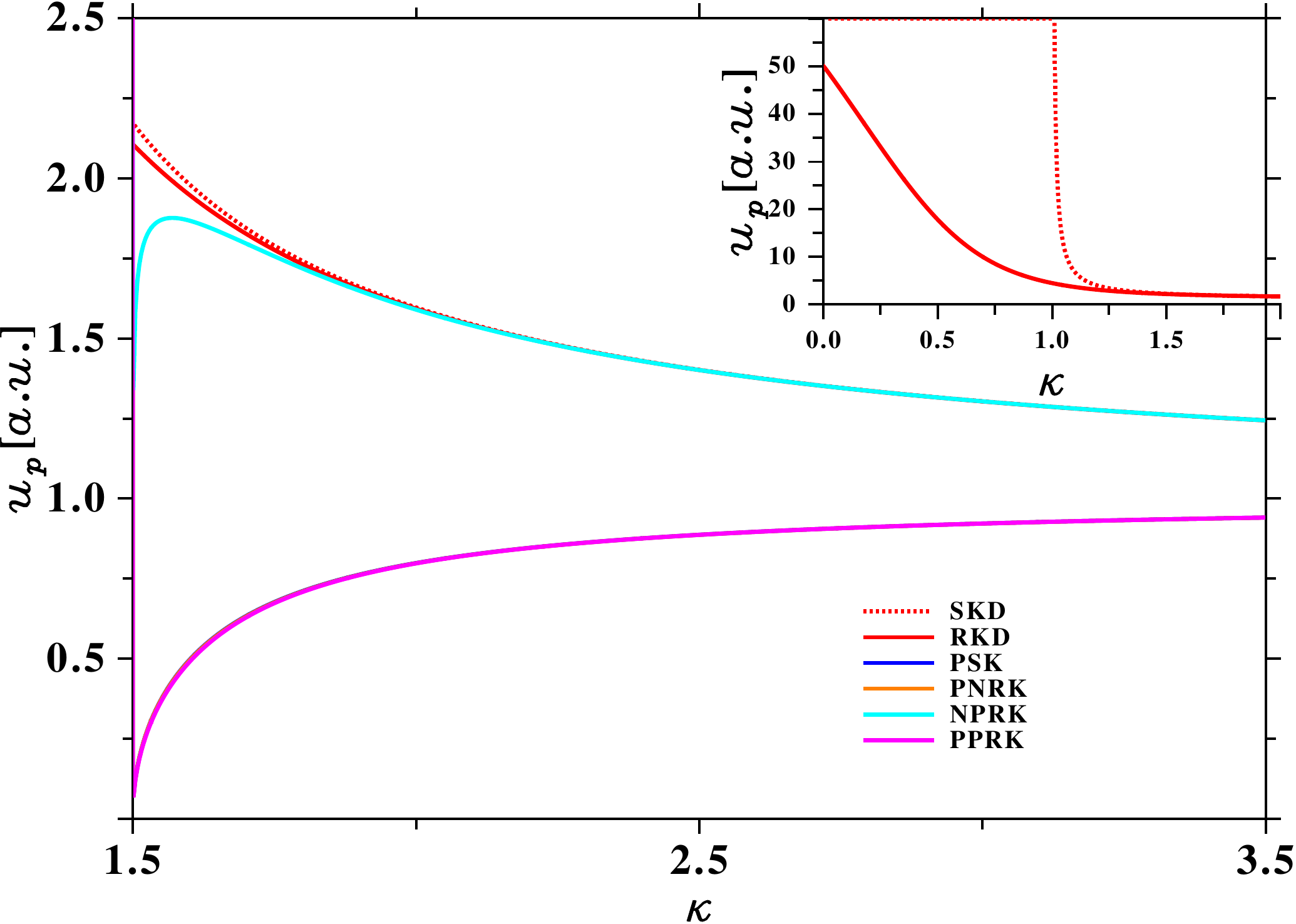}
  \caption{Most probable speed. The common factors given in
    Table~\ref{tab:3} are neglected, i.e.,\ $ \frac{2
      n_{0}}{\sqrt{\pi} }\Theta$. The left panel is for
    $\xi=0.1$ and the right panel for $\xi=0.01$. The inlay shows the values for
    $\kappa<2$ for the RKD and SKD. There it can nicely
    be seen that the lower limit for the RKD is $\frac{1}{2\xi}$. It
    can also be seen that the most probable speed for the SKD has its
    pole at $\kappa=1$ \citep{Scherer-etal-2019b}.\label{fig:1}}
\end{figure*}

\begin{figure*}
  \centering
  \includegraphics[width=0.45\textwidth]{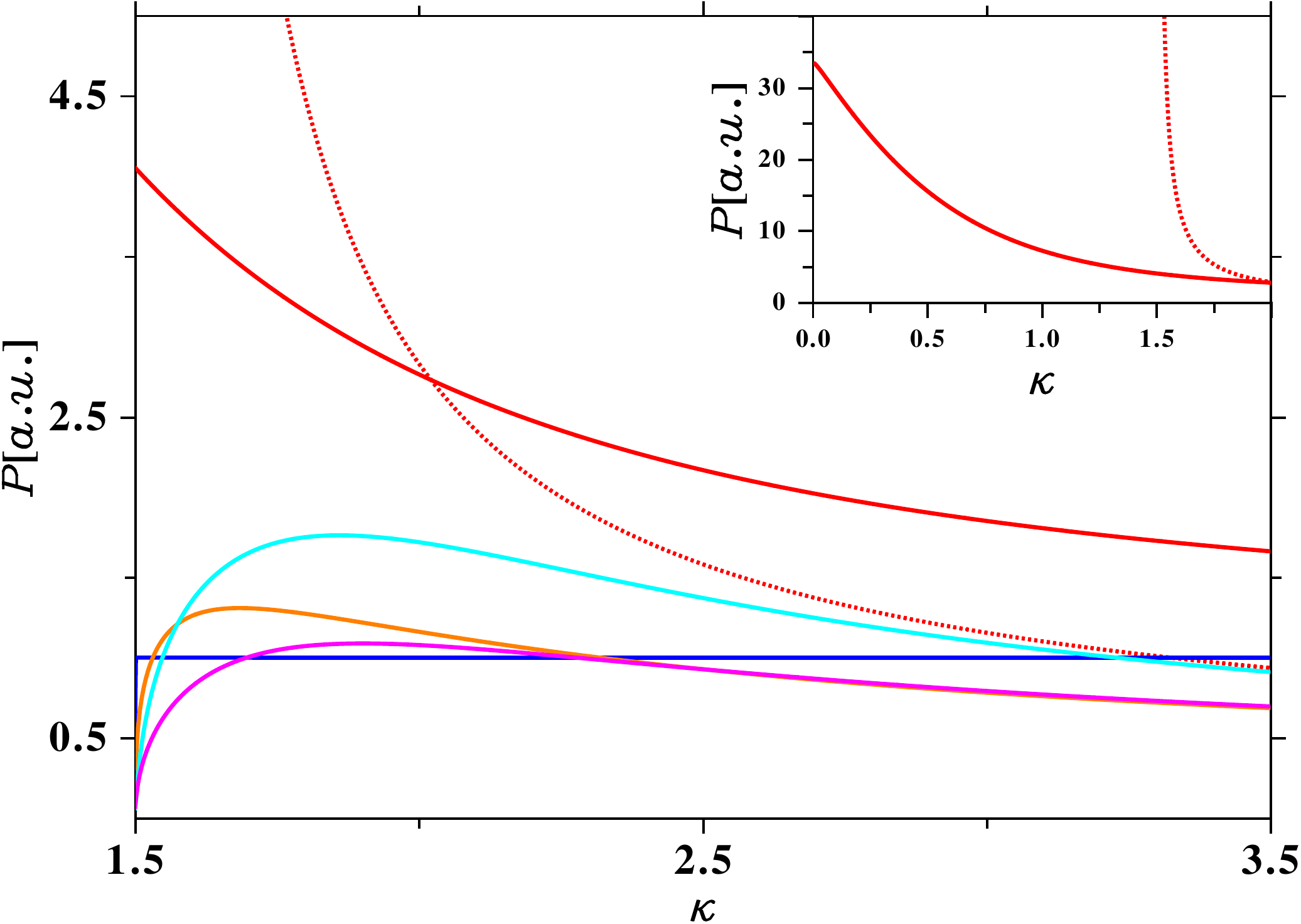}\hfill
  \includegraphics[width=0.45\textwidth]{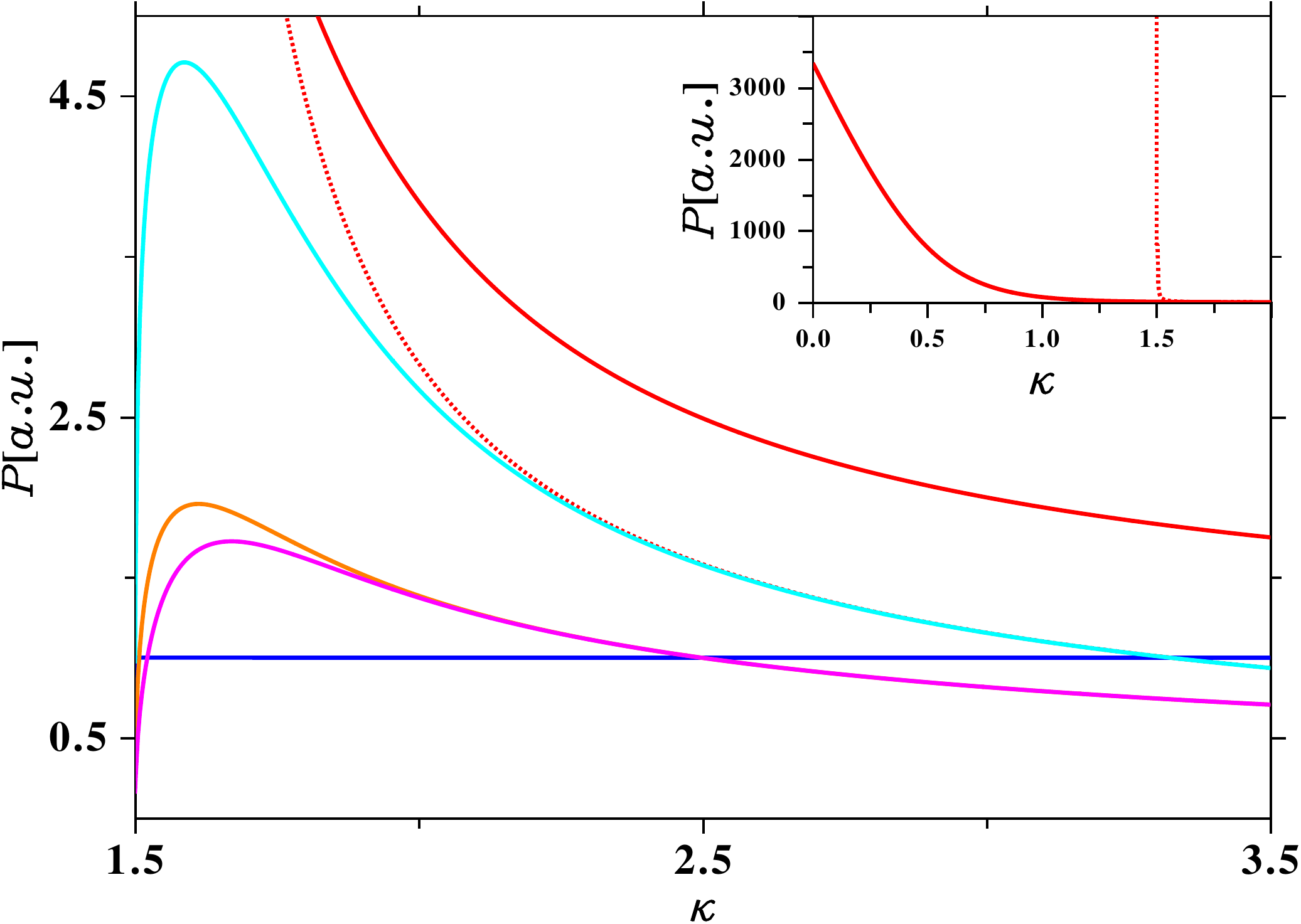}
  \caption{The pressure. The common factors given in
    Table~\ref{tab:3} are neglected, i.e.,\ $ \frac{2
      n_{0}}{3} \Theta^{2}$. The left panel is for
    $\xi=0.1$ and the right panel for $\xi=0.01$. The inlay shows the values for
    $\kappa<2$ for the RKD and SKD. There it can nicely
    be seen that the lower limit for the RKD is
    $\frac{1}{3\xi^{2}}$. \label{fig:2}}
\end{figure*}

In Table~\ref{tab:3a} the lower and upper limits are given (for the
\DL see the discussion in Section~\ref{sec:DL}). In
Fig.~\ref{fig:1} the non-Maxwellian part of the most probable speed
$\tilde{u}_{p}$
and in Fig.~\ref{fig:2} that of the pressure $\tilde{P}$
are plotted. From Table~\ref{tab:3} we see that the most probable
speed depends on either the square root of $\kappa$
or $\kappa-\thalf$,
while the pressure is linearly dependent on $\kappa$
or $\kappa-\thalf$.
Additionally, there are some factors which are specific to the
underlying distribution function: for the SKD these are
$\Gamma$-functions
$\left(\frac{1}{\kappa-\thalf}=\frac{\GA{\kappa-\thalf}}{\GA{\kappa-\half}}\right)$,
while for the distributions with an exponential part these are 
Kummer-$U$ functions. An
exception is the recipe $(\kappa-\thalf, \kappa+1, 0)$, where the most probable speed depends also on
$\Gamma$-functions,
the pressure is constant, i.e.,\ it does not depend on $\kappa$.

In Table~\ref{tab:3a} the lower and upper limits are given. One can see
that only for the SKD and RKD the most probable speed has finite
values when $\kappa$ reaches its limit $\kappa=\thalf$, for the
$\Psi$-functions the most probable speed vanishes. The upper limits
are for all distribution functions one or close to one.

A similar behavior can be seen for the pressure: The SKD goes to
infinity when $\kappa\to\thalf$, the RKD remains finite at a value
$1/(3\xi^{2})$, the recipe $(\kappa-\thalf, \kappa+1, 0)$ is constant, and the other three
distributions go to zero. The upper limits are again one or close to one

The limits are evaluated with the help of Table~\ref{tab:limit}
assuming that the second argument of the \KU, i.e., $\frac{\nu+3}{2}-\kappa$
is for low $\kappa$-values
in a very good approximation
$\lim\limits_{\kappa\to 0}\frac{\nu+3}{2}-\kappa = \frac{\nu+3}{2}$,
but the third argument $\kappa\xi^{2}$
is still not zero.  For $\kappa\to \infty$
it is easiest to go back to the original integral expression,  let
first $\kappa \to \infty$,
and than estimate the integrals. 

\begin{table*}
\begin{tabular}{l|L|KK|KK|KK}	
  \multicolumn{1}{c|}{name}
  &\multicolumn{1}{c|}{recipe}
  &\multicolumn{2}{c|}{limit $\tilde{u}_{p}$}
  &\multicolumn{2}{c|}{limit $\tilde{P}$}
  &\multicolumn{2}{c}{limit $\tilde{\Lambda}_{D}$}
  \\
  && \multicolumn{1}{c}{lower}  
  & \multicolumn{1}{c|}{upper}  
  & \multicolumn{1}{c}{lower}  
  & \multicolumn{1}{c|}{upper}  
  & \multicolumn{1}{c}{lower}  
  & \multicolumn{1}{c}{upper}
  \\
  \hline
  SKD & (\kappa,\kappa+1,0)
  & \lim\limits_{\kappa\to\frac{3}{2}} \tilde{u}_{p} = \frac{\sqrt{6\pi}}{2}\hfill~
  & \lim\limits_{\kappa\to\infty} \tilde{u}_{p} = 1\hfill~
  & \lim\limits_{\kappa\to\frac{3}{2}} \tilde{P}     = \infty\hfill~
  & \lim\limits_{\kappa\to\infty} \tilde{P} = 1\hfill~
  & \lim\limits_{\kappa\to\frac{3}{2}} \tilde{\Lambda}_{D}  = \frac{2}{3}\hfill~
  & \lim\limits_{\kappa\to\infty} \tilde{\Lambda}_{D} = 1  \hfill~    
  \\
  RKD &(\kappa,\kappa+1,\xi^{2})
  & \lim\limits_{\kappa\to 0} \tilde{u}_{p} =\frac{1}{2\xi}\hfill~
  & \lim\limits_{\kappa\to\infty} \tilde{u}_{p} = \frac{1}{\sqrt{1+\xi^{2}}}\hfill~
  & \lim\limits_{\kappa\to 0}  \tilde{P}=\frac{1}{3\xi^{2}}\hfill~
  & \lim\limits_{\kappa\to\infty} \tilde{P} = \frac{1}{1+\xi^{2}}\hfill~
  & \lim\limits_{\kappa\to 0} \tilde{\Lambda}_{D}= \frac{1}{1+\xi^{2}}\hfill~
  & \lim\limits_{\kappa\to\infty} \tilde{\Lambda}_{D} = \frac{1}{1+\xi^{2}}\hfill~
  \\
  \hline
  & (\kappa-\thalf, \kappa+1, 0) 
  & \lim\limits_{\kappa\to\frac{3}{2}} \tilde{u}_{p} = 0\hfill~
  & \lim\limits_{\kappa\to\infty} \tilde{u}_{p} = 1\hfill~
  & \lim\limits_{\kappa\to\frac{3}{2}} \tilde{P} = 1\hfill~
  & \lim\limits_{\kappa\to\infty} \tilde{P} = 1\hfill~
  & \lim\limits_{\kappa\to \thalf} \tilde{\Lambda}_{D}= 0\hfill~
  &\lim\limits_{\kappa\to\infty} \tilde{\Lambda}_{D} = 1\hfill~
   \\
  &(\kappa-\thalf, \kappa+1, \xi^{2}) 
  & \lim\limits_{\kappa\to\frac{3}{2}} \tilde{u}_{p} = 0\hfill~
  & \lim\limits_{\kappa\to\infty} \tilde{u}_{p} = \frac{1}{\sqrt{1+\xi^{2}}}\hfill~
  & \lim\limits_{\kappa\to\frac{3}{2}} \tilde{P} = 0\hfill~
  & \lim\limits_{\kappa\to\infty} \tilde{P} = \frac{1}{1+\xi^{2}}\hfill~
  & \lim\limits_{\kappa\to \thalf} \tilde{\Lambda}_{D}= 0\hfill~
  & \lim\limits_{\kappa\to\infty} \tilde{\Lambda}_{D} = \frac{1}{1+\xi^{2}}\hfill~
  \\
  &(\kappa, \kappa+1, \frac{\kappa\xi^{2}}{\kappa-\thalf}) 
  & \lim\limits_{\kappa\to\frac{3}{2}} \tilde{u}_{p} = 0\hfill~
  & \lim\limits_{\kappa\to\infty} \tilde{u}_{p} = 1\hfill~
  & \lim\limits_{\kappa\to\frac{3}{2}} \tilde{P} = 0\hfill~
  & \lim\limits_{\kappa\to\infty} \tilde{P} = 1\hfill~
  & \lim\limits_{\kappa\to \thalf} \tilde{\Lambda}_{D}= 0\hfill~
  & \lim\limits_{\kappa\to\infty} \tilde{\Lambda}_{D} = \infty\hfill~
   \\
  &(\kappa-\thalf, \kappa+1, \frac{\kappa\xi^{2}}{\kappa-\thalf}) 
  & \lim\limits_{\kappa\to\frac{3}{2}} \tilde{u}_{p} = 0\hfill~
  & \lim\limits_{\kappa\to\infty} \tilde{u}_{p} = 1\hfill~
  & \lim\limits_{\kappa\to\frac{3}{2}} \tilde{P} = 0\hfill~
  & \lim\limits_{\kappa\to\infty} \tilde{P} = 1\hfill~
  & \lim\limits_{\kappa\to \thalf} \tilde{\Lambda}_{D}= 0\hfill~
  & \lim\limits_{\kappa\to\infty} \tilde{\Lambda}_{D} = 1\hfill~
\end{tabular}
\caption{\label{tab:3a}The limits for the most probable speed,
  pressure and \DL{s}. \label{tab:dl}}
\end{table*}

From Table~\ref{tab:3} and from Figs.~\ref{fig:1} and~\ref{fig:2} one
can see that the most probable speed and pressures for the SKD and RKD
behave as expected: The most probable speeds are proportional to
$\sqrt{\kappa}$ (and to those in the $\Gamma$-function)
and are defined for the SKD for all $\kappa\ge \thalf$
and for the RKD for all $\kappa\ge 0$.
In both cases the most probable speed is monotonically decreasing with
increasing $\kappa$, see Fig.~\ref{fig:1}. The same holds true for the
pressure, see Fig.~\ref{fig:2}

For the $\Psi$-distributions
all speeds go to zero when $\kappa\to\thalf$.
The recipe $(\kappa, \kappa+1, \frac{\kappa\xi^{2}}{\kappa-\thalf})$ distribution shows the best approximation to the SKD or RKD
for higher $\kappa$-values,
while the other $\Psi$-distributions
(the recipes $(\kappa-\thalf, \kappa+1, 0)$,  $(\kappa-\thalf, \kappa+1, \xi^{2})$ and  $(\kappa-\thalf, \kappa+1, \frac{\kappa\xi^{2}}{\kappa-\thalf})$) lie ``much'' below the reference curves of
the SKD or RKD.

A similar behavior for the pressure can be seen in Fig.~\ref{fig:2}: by
definition the recipe $(\kappa-\thalf, \kappa+1, 0)$ pressure is constant (equal to the Maxwellian one),
but for the recipes $(\kappa-\thalf, \kappa+1, \xi^{2})$,  $(\kappa, \kappa+1, \frac{\kappa\xi^{2}}{\kappa-\thalf})$ and  $(\kappa-\thalf, \kappa+1, \frac{\kappa\xi^{2}}{\kappa-\thalf})$ distributions the pressure goes to
zero when $\kappa\to\thalf$.
The SKD goes to infinity for $\kappa\to\thalf$,
while the RKD has a finite value for $\kappa > 0$ depending also on $\xi$.

The higher-order moments behave similar (not shown): Except for the
recipe $(\kappa-\thalf, \kappa+1, 0)$, the higher-order moments $M_{n>2}$
for the $\Psi$-distributions
approach zero with decreasing $\kappa$.
The higher-order moments of the recipe $(\kappa-\thalf, \kappa+1, 0)$ behave similar to the SKD and require higher
$\kappa$-values to be defined ($\kappa\ge \frac{n+1}{2}$).

\section{The Debye length}\label{sec:DL}

The \DL is discussed in \citet{Treumann-etal-2004}, \citet{Livadiotis-McComas-2014} and
\citet{Fahr-Heyl-2016} as well as in \citet[][and references therein]{Livadiotis-etal-2018}. These
authors use the Debye-H\"uckel theory to determine the \DL 
(see Appendix~\ref{Debye}). Because we
are mainly dealing with collisionless plasmas, we apply here the
approach discussed in \citep{Krall-Trivelpiece-1973}. 
In this approach a uniformly moving point test charge is considered to cause a small perturbation 
    in a Vlasov plasma, which is otherwise uniform and field-free. By linearizing the plasma distribution 
    function and performing a Fourier-Laplace transformation of the Vlasov equation, the potential in the 
    plasma created by the test charge can be calculated, from which then the Debye length can be derived 
    (see Appendix~\ref{subsec:krall_trivelpiece}).
For distributions, which have only a single factor
depending on $v$ like the Maxwellian, SKD, and recipe $(\kappa-\thalf, \kappa+1, 0)$, we obtain the same
results in both approaches. But when we have two or more factors
depending on $v$, like all the distributions with a cutoff, we get
slightly different results.

We have derived the \DL $\Lambda$
for a single species in the appendix with a Maxwellian normalization factor
\begin{align}
  \Lambda_{M}^{2} = 2\frac{\epsilon_{0}m_{s}\Theta_{s}^{2}}{q^{2}n_{s}}\,,
\end{align}
which are given in Table~\ref{tab:3} and their limits in
Table~\ref{tab:3a} (to save writing we have dropped the index $s$ in
what follows).
\label{fig:U}
It can be seen from Table~\ref{tab:3a} that the lower limits $\kappa\to\thalf$ of the
$\Psi$-distributions for the \DL are always zero. In that case we
do not have a plasma, because the plasma parameter
$N_{p}=\frac{4\pi}{3}n_{0}\Lambda^{3}$ is then zero. For $\kappa$-values
close to the limit that may be also the case for the RKD, which has
always a finite \DL, but the plasma parameter may be
low. For $\kappa\to\infty$ the discussed \DL approaches
that of the Maxwellian \DL.

\begin{figure}
  \centering
  \includegraphics[width=0.45\textwidth]{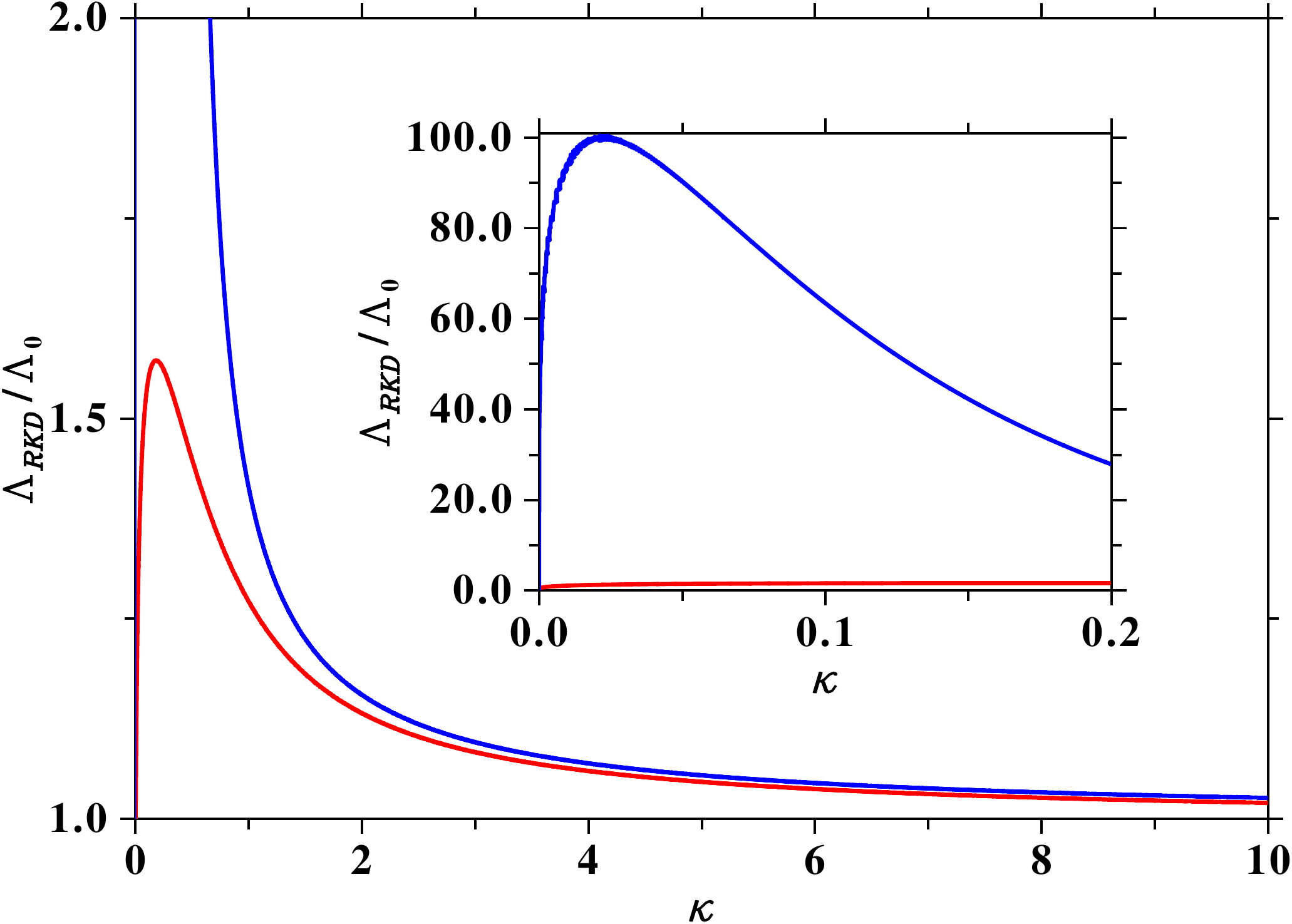}
  \caption{The \DL $\Lambda_{RKD}$ of the RKD  normalized to
    that of the Maxwellian $\Lambda_{M}$. The red curves represent the
  case $\xi=0.1$, while the blue curve shows $\xi=10^{-5}$. 
  It can be
  seen that both curves go in the limit $\kappa\to0$ to
  $\frac{1}{1-\xi^{2}}$ and for $\kappa\to\infty$ to the same
  limit. The maximum is at very low $\kappa$-values, and can be
  huge. Curves for $10^{-5}\le\xi\le 10^{-1}$ lie between the
  above two limiting curves. It can be seen that at the maximum for
  $\xi=10^{-5}$ the numerics are not very good. The inlet enlarges
  the view for low $\kappa$ values.\label{fig:dm}}
\end{figure}

\begin{figure}
  \centering
  \includegraphics[width=0.45\textwidth]{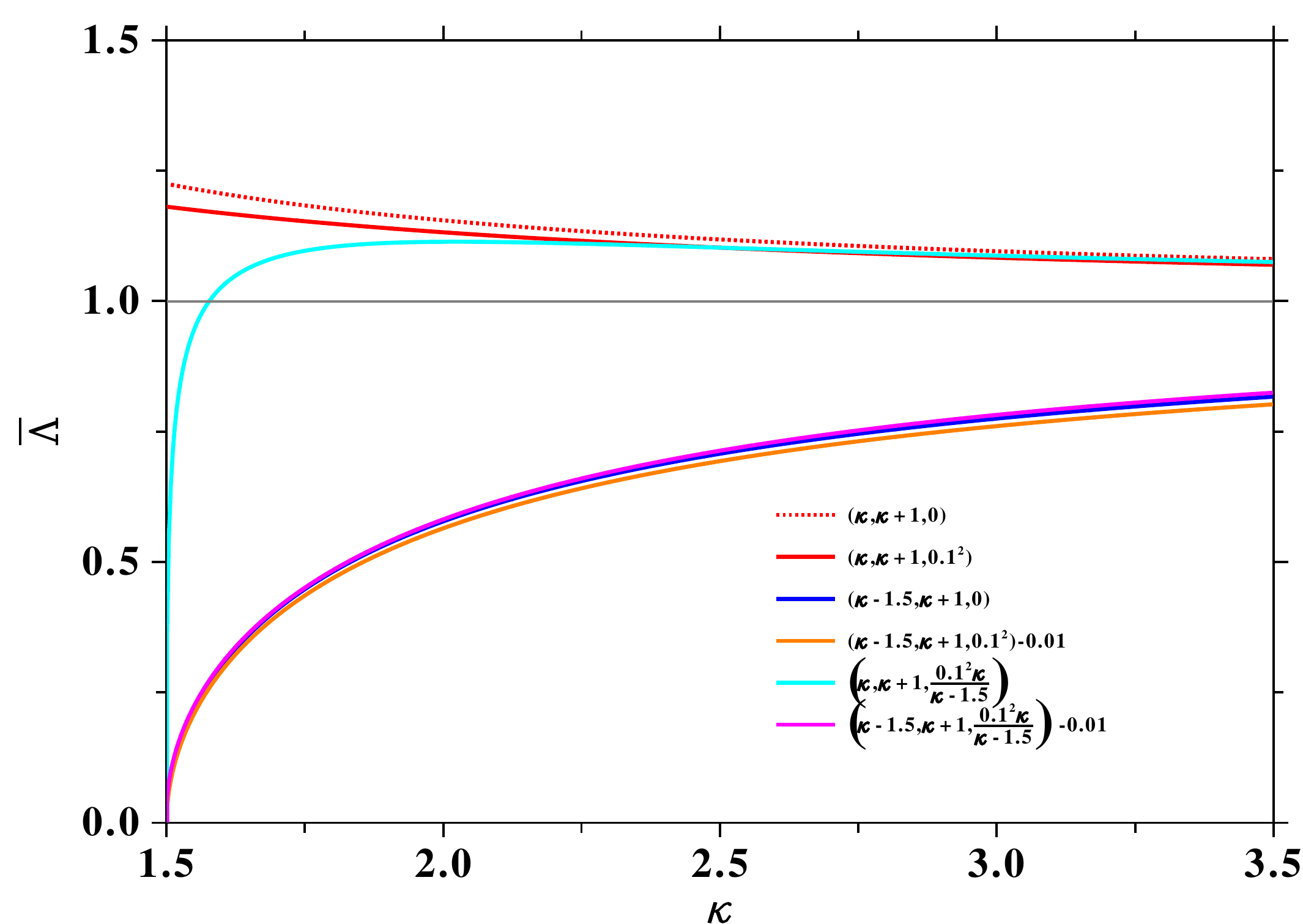}
  \caption{\label{fig:pk}The \DL for the $\Psi$-distributions
    and the SKD. It is obvious, that the $\Psi$-distributions, except
    the recipe $(\kappa, \kappa+1, \frac{\kappa\xi^{2}}{\kappa-\thalf})$, approach the upper limit from below, while the SKD from above.  }
\end{figure}

The \DL for the RKD is shown in Fig.~\ref{fig:dm} for two
$\xi$-values
and those for the SKD and $\Psi$-distributions
in Fig.~\ref{fig:pk}. The latter are shifted by a small amount (see
legend), because otherwise they would lie more or less on top of each other.
For the RKD is can be seen from Fig.~\ref{fig:dm} that for all
$\xi$-values the lower limit is close to that of the Maxwellian
\DL, which is also the case for the upper limit. In between,
at low $\kappa$-values, the RKD has a maximum, which can be quite
high for low $\xi$-values.  The RKD and the SKD
(Fig.~\ref{fig:pk}) approach the Maxwellian limit from above, while
the $\Psi$-distributions (except the recipe $(\kappa, \kappa+1, \frac{\kappa\xi^{2}}{\kappa-\thalf})$) approach them from below.
In all cases except that of recipe $(\kappa, \kappa+1, \frac{\kappa\xi^{2}}{\kappa-\thalf})$, if one is not too close to
the lower $\kappa$ limit,
the Debye radius is close to that of the Maxwellian
distribution. Furthermore, because the upper and lower limit for the
RKD are equal, the expected extremum (maximum) can clearly be
seen. For low $\xi$-values
it can be quite high, and drops quickly to the limit, when
$\kappa\to 0$.
The limit for $\kappa\to\infty$
is not reached for the $\kappa$-values shown in Figs.~\ref{fig:dm} and~\ref{fig:pk}.

Note: The \DL $\Lambda$
is a combination of the Debye lengths of all species. Thus, following
Eq.~(\ref{eq:dl}) we can also have the case that one species (say the electrons)
is, for example, $\kappa$-distributed, while the other species (the ions) is Maxwellian
distributed. Even if all species are $\kappa$-distributed 
it is very unlikely that both have the same $\kappa$-value. 
Moreover, by definition of a plasma the number of particles in
a Debye sphere must be large, which is not the case when $\Lambda\to 0$,
except when simultaneously the number density goes to infinity.

\section{Application to observations}\label{sec:appl}

\subsection{Pressure and heat flow for a sum of distribution functions}
To estimate the pressure and heat flow for
  the sum of distribution functions \citep{Paschmann-etal-1998} is more complicated: 
  It has to be taken in the center of mass system, which can be calculated
  as the first moment $n\vec{u}$:
\begin{align}
    \vec{u} = \frac{\sum\limits_{i=1}^{i_{max}}m_{i}n_{i}\vec{u}_{i}}
    {\sum\limits_{i=1}^{i_{max}}m_{i}n_{i}}\,,
\end{align}
where $\vec{u}_{i}$ is is the drift velocity of the species $i$, 
$n_{i}$ its number density and $m_{i}$ its mass. With $i_{max}$ the total
number of distribution functions involved is given.   
Thus the pressure is
\begin{align}\label{eq:pressure}
  \tensor{P} &= \sum\limits_{i=1}^{i_{max}} m_{i}\int
  f_{i}(\vec{v}-\vec{u_{i}})
               (\vec{v}-\vec{u})\otimes(\vec{v}-\vec{u})\md^{3}v \\\nonumber
  &=
  \sum\limits_{i=1}^{i_{max}} m_{i}\int
  f_{i}(\vec{v})
  (\vec{v}-\vec{u}+\vec{u}_{i})\otimes(\vec{v}-\vec{u}+\vec{u}_{i})\md^{3}v\\\nonumber
    & = \sum\limits_{i=1}^{i_{max}} m_{i}\int f_{i}(\vec{v})
      \left[\underbrace{\vec{v}\otimes\vec{v}}_{\substack{\mathrm{partial\
      thermal}\\
      \mathrm{pressure}\tensor{P}_{i}}}
     - \underbrace{(\vec{u}-\vec{u}_{i})\otimes(\vec{u}-\vec{u}_{i})}
      _{\mathrm{ internal\ ram\ pressure\ }\tensor{R}_{i}}
      \right] \md^{3}v\,,
\end{align}
were we have neglected all the terms of the odd functional
dependencies of $\vec{v}$ 
and the double sided arrow over quantities denotes second-order tensors. The terms 
    in the square brackets of Eq.~\eqref{eq:pressure} can be identified as the 
    (isotropic) partial pressure tensor $\tensor{P}_{i}$ due to thermal motion, and the internal 
    ram pressure tensor $\tensor{R}_{i}$ due to the bulk motion, each contributed by particle species $i$.

For isotropic distributions we are only interested in the trace of the
pressure tensor ($\vec{u}$ and $\vec{u_{i}}$  (in Cartesian coordinates)):
\begin{align}\nonumber
  \tr{P} &= \sum\limits_{i=1}^{i_{max}} m_{i}
           \left(\tr{P_{i}} - \tr{R_{i}}\right) \\
  &=
  \sum\limits_{i=1}^{i_{max}} m_{i}\left(3 P_{11,i} - n_{i} (\vec{u}-\vec{u}_{i})^{2}\right)\,.
\end{align}
For the heat flux the expression is much more
complicated\footnote{Eq.~(\ref{hcorr}) is valid also for asymmetric
  distribution functions. In  Eq.s~(11) and~(12)
  (as well as in the appendix Eq.s~(B17) and~(B20)) given by
  \citet{Scherer-etal-2019a} there are factors 2 missing.}:
\begin{align}\nonumber
  \vec{H} &= \half \sum\limits_{i=1}^{i_{max}} m_{i}\int
  f_{i}(\vec{v}-\vec{u_{i}})(\vec{v}-\vec{u})^{2}(\vec{v}-\vec{u})\md^{3}v\\\nonumber
            &=\half \sum\limits_{i=1}^{i_{max}} m_{i}\int
            f_{i}(\vec{v})(\vec{v}-\vec{u}+\vec{u}_{i})^{2}(\vec{v}-\vec{u}+\vec{u}_{i})\md^{3}v\\\nonumber
          & = \half \sum\limits_{i=1}^{i_{max}} m_{i}\int f_{i}
            \left[v^{2}\vec{v} - 2 (\vec{v}\cdot\vec{U}_{i})\vec{v} +
            U_{i}^{2}\vec{v} - v^{2}\vec{U}_{i} +
            2(\vec{v}\cdot\vec{U}_{i})\vec{U}_{i} -
            U_{i}^{2}\vec{U}_{i}\right]\\\label{hcorr}
  &= \half \sum\limits_{i=1}^{i_{max}} m_{i}\left(2\tensor{P_{i}} + \tr{P_{i}} - n_{i}U_{i}^{2}\right)\vec{U_{i}}
\end{align}
with $\vec{U}_{i}=\vec{u}-\vec{u}_{i}$. 
The above pressure will be used in the following.

\subsection{Fit to observations}
%
\subsubsection{High $\kappa$-values}
\begin{figure}
   \includegraphics[width=0.45\textwidth]{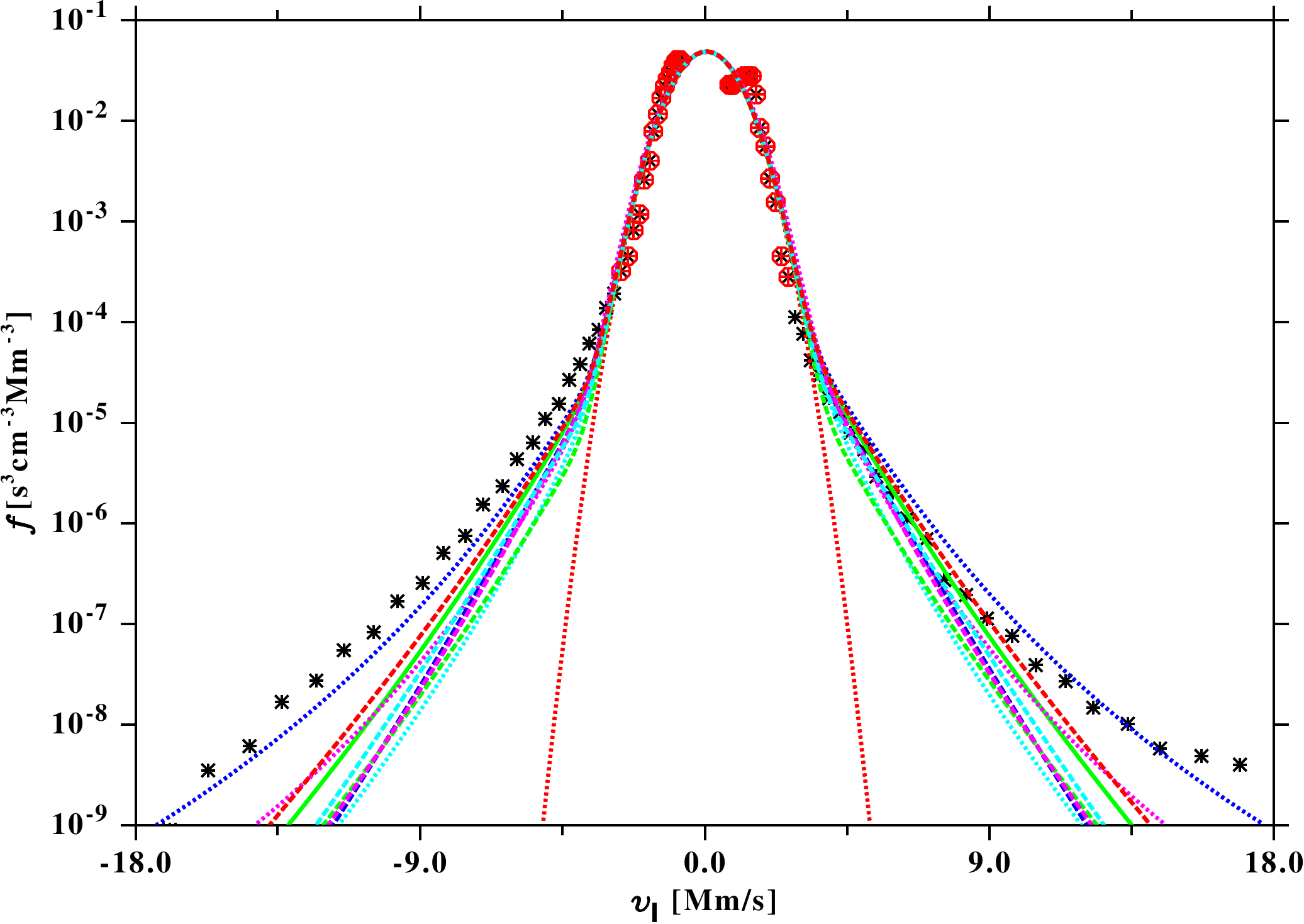}\hfill
   \includegraphics[width=0.45\textwidth]{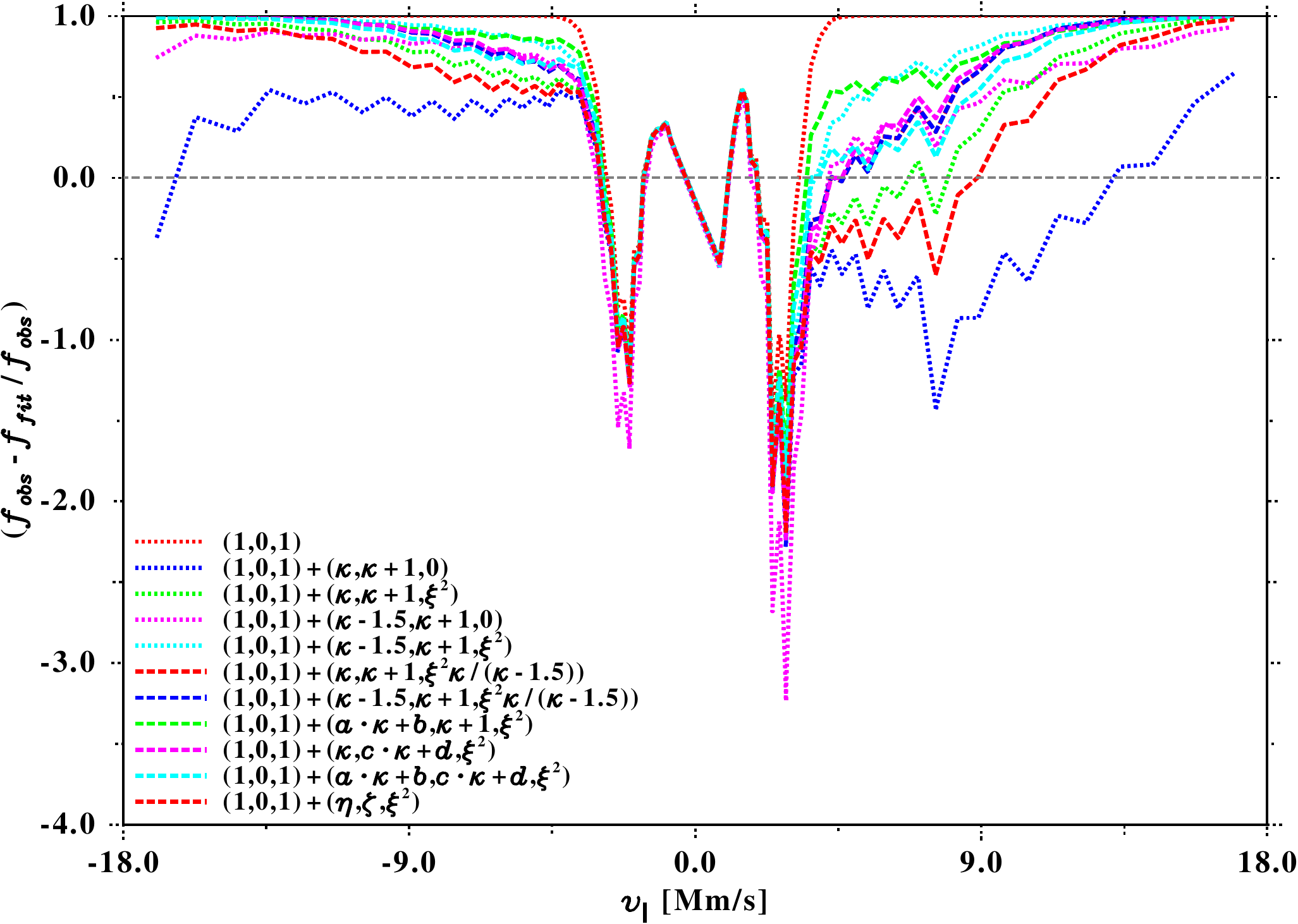}
  \caption{Top panel: The stars mark the Ulysses observations of the parallel
    part of an electron distribution. The recipes fitted are given
    in the legend. Bottom panel: The relative errors of recipes. See main
    text for more information.\label{fig:Ulysses}}
\end{figure}

We demonstrate the use of the above discussed recipes using the electron data set from 
Ulysses, the event from 15.01.2002 at 3:33:42, which are shown in Fig.~\ref{fig:Ulysses} 
for the parallel velocity component, i.e., only those parallel to the magnetic field. 
First, we assume that this is an ideal observed distribution function with no error bars. 
As discussed in \citet{Lazar-etal-2017}, see also \citet{Maksimovic-etal-2005} and \citet{ 
Stverak-etal-2008}, these distributions $f$ are best fitted by composite model distributions 
$f(v)= f_c(v) + f_h(v) + f_s(v)$, combining a quasithermal core (subscript $c$) at 
low-energies, well reproduced by a standard Maxwellian, and two suprathermal components, 
the central halo (subscript $h$) and the field-aligned strahl (subscript $s$), each of them 
best fitted by Kappa power-laws. We may therefore reduce to a dual model
 \begin{align}
   f = f_{M}(v,\Theta_{M},u_{M}) + f_{\kappa}(v,\Theta_{\kappa},u_{\kappa})\,,
 \end{align}
where $f_{M}$ is the Maxwellian core and $f_{\kappa}$ incorporates the suprathermal (halo 
and strahl) populations. The recipes used for $f_{\kappa}$ are indicated in the legends of
Fig.~\ref{fig:Ulysses}. In a referential fixed to protons distribution functions may be
assumed to depend not only on the velocity $v$ and the core speed $\Theta$, but also on a
drift (bulk) speed $u$. The non-linear fit is done in the following way: First we fit the 
Maxwellian part using only the values, which are in the 3~$\sigma$ of the maximum value 
in the data set. These
points are indicated by the red stars in the left panel of Fig.~\ref{fig:Ulysses}. The 
remaining data set is indicated by the black stars. In the next step we use these values as 
the starting point for the Maxwellian core in the combined distribution function, and then 
fit suprathermal tails, first with a SKD. The obtained fitted values are then used as 
reference for all the other recipes.
 
 \begin{table*}
 \scalebox{0.7}{
 \begin{tabular}{l|lll|lllllc|llll}
 Recipe & $n_{M}$ [cm$^{-3}$] & $\Theta_{M}$ [Mm$^{-3}]$& $u_{M}$  [Mm$^{-3}]$& 
 $n_{\kappa}$[cm$^{-3}]$ & $\Theta_{\kappa}$ [Mm$^{-3}]$ & $u_{\kappa}$ [Mm$^{-3
 }]$ & $\eta$& $\zeta$ & $\xi$ & $a$ & $b$ & $c$ & $d$\\ 
 &\multicolumn{3}{c|}{~} & \multicolumn{3}{c}{~}&\multicolumn{2}{c}{$\kappa$} &$
 \xi$ &&&&\\
 \hline
$(1,0,1)$                                                                                            &      0.505 &     1.227 &     0.054 &
 &&&&&&&&\\
$(1,0,1)+(\kappa ,\kappa+1,0)$                                                                       &      0.505 &     1.238 &     0.054 &     0.027 &     2.180 &     0.182  &\multicolumn{2}{c}{     3.267 } & 0.000E+00  &
 &&&\\
$(1,0,1)+(\kappa ,\kappa+1,\xi^2)$                                                                   &      0.507 &     1.234 &     0.054 &     0.025 &     2.159 &     0.167  &\multicolumn{2}{c}{     3.411 } & 0.523E-01  &
 &&&\\
$(1,0,1)+(\kappa-1.5,\kappa+1,0)$                                                                    &      0.553 &     1.278 &     0.060 &     0.027 &     2.282 &     0.178  &\multicolumn{2}{c}{     3.473 } & 0.565E-01  &
 &&&\\
$(1,0,1)+(\kappa-1.5,\kappa+1,\xi^2)$                                                                &      0.507 &     1.236 &     0.054 &     0.025 &     2.162 &     0.168  &\multicolumn{2}{c}{     3.412 } & 0.524E-01  &
 &&&\\
$(1,0,1)+(\kappa ,\kappa+1,\xi^2 \kappa /(\kappa-1.5))$                                              &      0.507 &     1.236 &     0.054 &     0.025 &     2.161 &     0.168  &\multicolumn{2}{c}{     3.413 } & 0.523E-01  &
 &&&\\
$(1,0,1)+(\kappa-1.5 ,\kappa+1,\xi^2\kappa /(\kappa-1.5))$                                           &      0.507 &     1.236 &     0.054 &     0.025 &     2.160 &     0.168  &\multicolumn{2}{c}{     3.414 } & 0.523E-01  &
 &&&\\
$(1,0,1)+(a\cdot\kappa+b ,\kappa+1,\xi^2)$                                                           &      0.508 &     1.235 &     0.054 &     0.025 &     2.162 &     0.168  &\multicolumn{2}{c}{     3.418 } & 0.524E-01  &     1.677 &     1.501 &     0.000 &     0.000 
 \\
$(1,0,1)+(\kappa ,c\cdot\kappa+d,\xi^2)$                                                             &      0.507 &     1.236 &     0.054 &     0.025 &     2.159 &     0.168  &\multicolumn{2}{c}{     3.417 } & 0.524E-01  &  &  &     1.015 &     1.549 
 \\
$(1,0,1)+(a\cdot\kappa+b ,c\cdot\kappa+d,\xi^2)$                                                     &      0.507 &     1.235 &     0.054 &     0.025 &     2.161 &     0.168  &\multicolumn{2}{c}{     3.417 } & 0.523E-01  &     1.601 &     0.850 &     1.375 &     0.866 
 \\
$(1,0,1)+(\eta , \zeta ,\xi^2)$                                                                      &      0.508 &     1.235 &     0.054 &     0.025 &     2.159 &     0.168  &     3.414  &     4.160  & 0.524E-01  &
 &&&\\
 \hline
 $<\bullet>$ &
     0.511 &     1.239 &     0.055 &     0.026 &     2.176 &     0.170 &\multicolumn{2}{c}{     3.405 } & 0.470E-01 &
 &&&\\
 $\sigma(\bullet)$ &
 0.140E-01 & 0.131E-01 & 0.160E-02 & 0.811E-02 & 0.689E+00 &     0.054 &\multicolumn{2}{c}{     1.078 } & 0.217E-01 &
 &&&
 \end{tabular}
 }
 \caption{\label{tab:fit} The number densities $n_{M},n_{\kappa}$, core speeds
 $\Theta_{M}, \Theta_{\kappa}$
 and drift speeds $u_{M},u_{\kappa}$ for the Maxwellian part $M$ and the
 $\kappa$ recipies. 
 The latter are given in the first column. The last two rows give the mean values and
 standard deviation
 for the corresponding column. See text for further discussion}
 \end{table*}
 \begin{table*}
 \scalebox{0.7}{
 \begin{tabular}{l|lll|lll}
 Recipe & $u_{p,M}$ [km/s] & $P_{M} $ [m$^{-1}$s$^{-1}$]& $R_{M}$ [m$^{-1}$s$^{-
 1}$] & $u_{p,\kappa}$[km/s] &  $P_{\kappa} $ [m$^{-1}$s$^{-1}$]& $R_{\kappa}$ [
 m$^{-1}$s$^{-1}$]\\
 \hline
$(1,0,1)$ &      699.4 &     0.507 &    0.0015 &       0.0 &     0.000 &    0.0000 \\$(1,0,1)+(\kappa ,\kappa+1,0)$ &      706.3 &     0.517 &    0.0015 &      84.9 &     0.159 &    0.0009 \\$(1,0,1)+(\kappa ,\kappa+1,\xi^2)$ &      705.5 &     0.514 &    0.0015 &      70.4 &     0.085 &    0.0007 \\$(1,0,1)+(\kappa-1.5,\kappa+1,0)$ &      797.9 &     0.602 &    0.0020 &      65.3 &     0.094 &    0.0009 \\$(1,0,1)+(\kappa-1.5,\kappa+1,\xi^2)$ &      706.8 &     0.516 &    0.0015 &       0.0 &     0.000 &    0.0007 \\$(1,0,1)+(\kappa ,\kappa+1,\xi^2 \kappa /(\kappa-1.5))$ &      707.4 &     0.517 &    0.0015 &       0.0 &     0.000 &    0.0007 \\$(1,0,1)+(\kappa-1.5 ,\kappa+1,\xi^2\kappa /(\kappa-1.5))$ &      707.1 &     0.516 &    0.0015 &       0.0 &     0.000 &    0.0007 \\$(1,0,1)+(a\cdot\kappa+b ,\kappa+1,\xi^2)$ &      707.4 &     0.516 &    0.0015 &      66.0 &     0.065 &    0.0007 \\$(1,0,1)+(\kappa ,c\cdot\kappa+d,\xi^2)$ &      707.7 &     0.517 &    0.0015 &      63.9 &     0.073 &    0.0007 \\$(1,0,1)+(a\cdot\kappa+b ,c\cdot\kappa+d,\xi^2)$ &      707.2 &     0.516 &    0.0015 &      57.3 &     0.055 &    0.0007 \\$(1,0,1)+(\eta , \zeta ,\xi^2)$ &      707.1 &     0.516 &    0.0015 &      73.8 &     0.091 &    0.0007 \\
 \hline
 $<\bullet>$ &
     714.5 &     0.523 &    0.0015 &      43.8 &     0.057 &    0.0007 \\
 $\sigma(\bullet)$ &
      26.4 &     0.025 &    0.0001 &      33.7 &     0.050 &    0.0002
 \end{tabular}
 }
 \caption{\label{tab:presfit} For the above recipes (also first column)
 the most probable speed, the thermal pressure and the ram pressure are
 presented. To get physical units, the pressure terms must be multiplied by 
 the mass. The last two rows give the mean values and standard deviations for
 the corresponding column. See text for further information.
 }
 \end{table*}

 In the upper top panel of Fig.~\ref{fig:Ulysses} the recipes are fitted.  It can be seen that the
   fit approximates the right flank of the data better, compared to
   the left flank.
 That hints to an asymmetry or to the missing (explicit) fit for an
 additional "strahl"-component.  Nevertheless, all discussed recipes
 lead to similar results concerning the fitted parameters,
 which are given in Table~\ref{tab:fit}. In the last two rows of this
 table we present the mean value and the variance of corresponding
 parameters given by the discussed recipes.
 Although fits of some recipes may show deviations from the
   observations,
 it turns out that all the fitted functions lead to quite similar
 results with a quite small variance.

 We can compare with plasma parameters provided by complementary
 measurements\footnote{taken from
   \verb+https://omniweb.gsfc.nasa.gov/+}
 at the time of observation of the above distribution function, e.g.,
 the average solar wind speed was $u_{sw}=636.3$\,km/s
 with a temperature of $T_{sw}=104,599$\,K
 and a proton number density of $n_{p}=0.22$\,cm$^{-3}$
 and a magnetic field strength of $B=0.75$\,nT.
 The latter gives an Alvf\'en speed $v_{A}= 35.6$\,km/s.

Thus, the electron number density is by a factor 2 too high to
guarantee charge neutrality. This is caused because we fit a spherical
distribution function instead an anisotropic one to the data. If we
assume that the maximum values of both distribution functions are
equal, we find that the perpendicular must be of the order of
0.66\,Mm. The average thermal speed of the core (Maxwellian) component $\Theta_{M}$ is 
in a very good agreement with the electron sound speed $c_{e}=1.266$\,Mm/s, while the 
average drift speed of the core is about the Alfv\'en speed. For the suprathermal component 
in the distribution function the average number density is roughly an order of magnitude 
less than the corresponding Maxwellian, but the average thermal speed is above twice that 
of Maxwellian core, and the average drift speed is three times higher than 
the core drift (see Table~\ref{tab:fit}).
Beside that the fits of the different recipes may not be that good, the
above results show that all the discussed distribution functions lead
to very similar macroscopic parameters like density, thermal and drift
speeds.

The next criteria are the physical parameters, namely, the most probable speeds $u_{M}$
or $u_{\kappa}$, the thermal pressures $P_{M}$ and $P_{\kappa}$ and the ram pressure $R_{M}$
and $R_{\kappa}$. These values are calculated using Eq.~(\ref{eq:M_cook}). The results are
presented in Tab~\ref{tab:presfit}. These parameters are also quite independent of the 
choice of the recipe. Also, the $\kappa$-values are all in the order of $\kappa=3$.
Thus, there is also no reason to discard one or the other discussed recipe, 
though the RKD-like models may present indubitable advantages \citep{Scherer-etal-2017,
Scherer-etal-2019a, Scherer-etal-2019b}.

\subsubsection{Low $\kappa$-values}
\begin{figure}
   \includegraphics[width=0.45\textwidth]{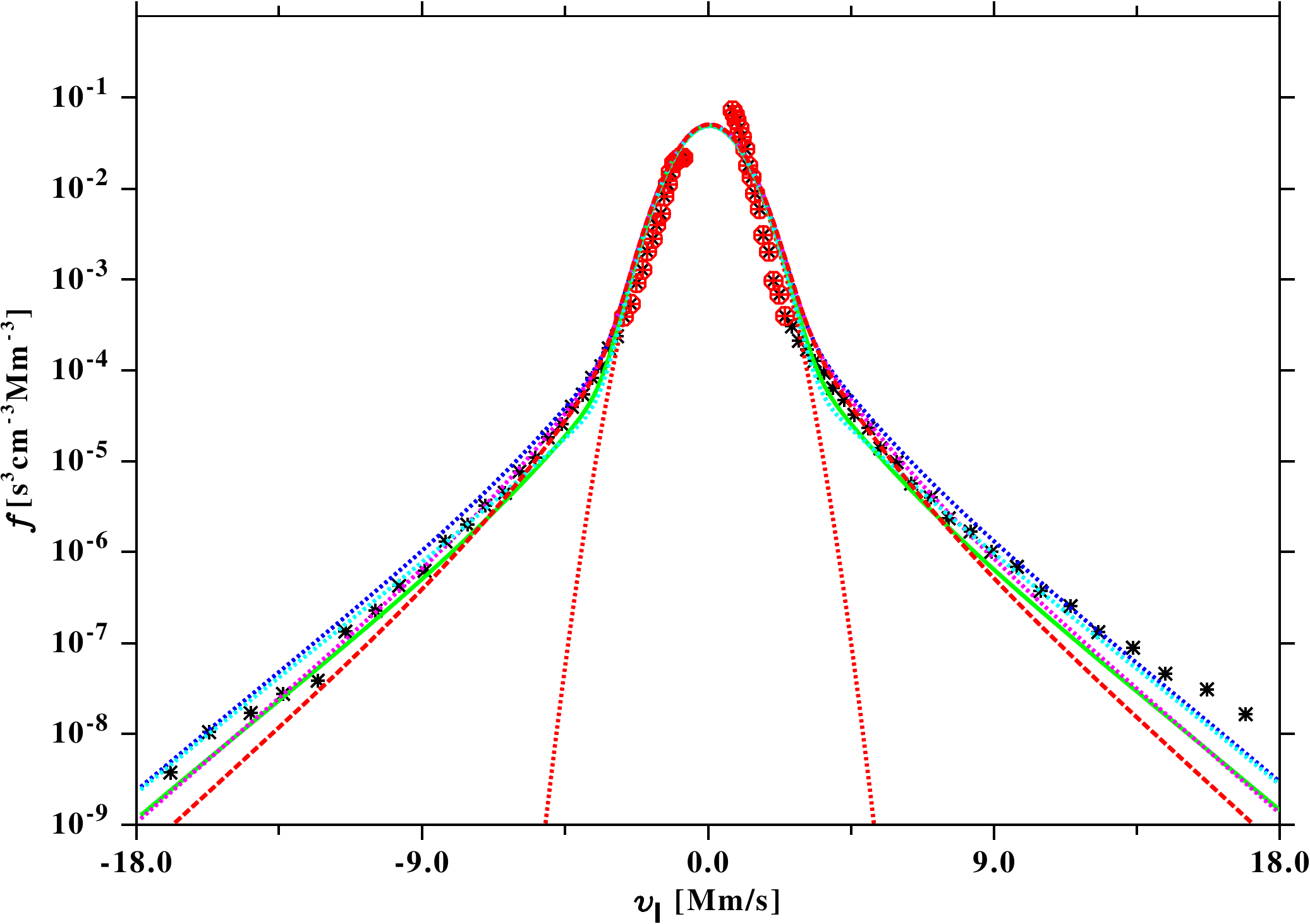}\hfill
   \includegraphics[width=0.45\textwidth]{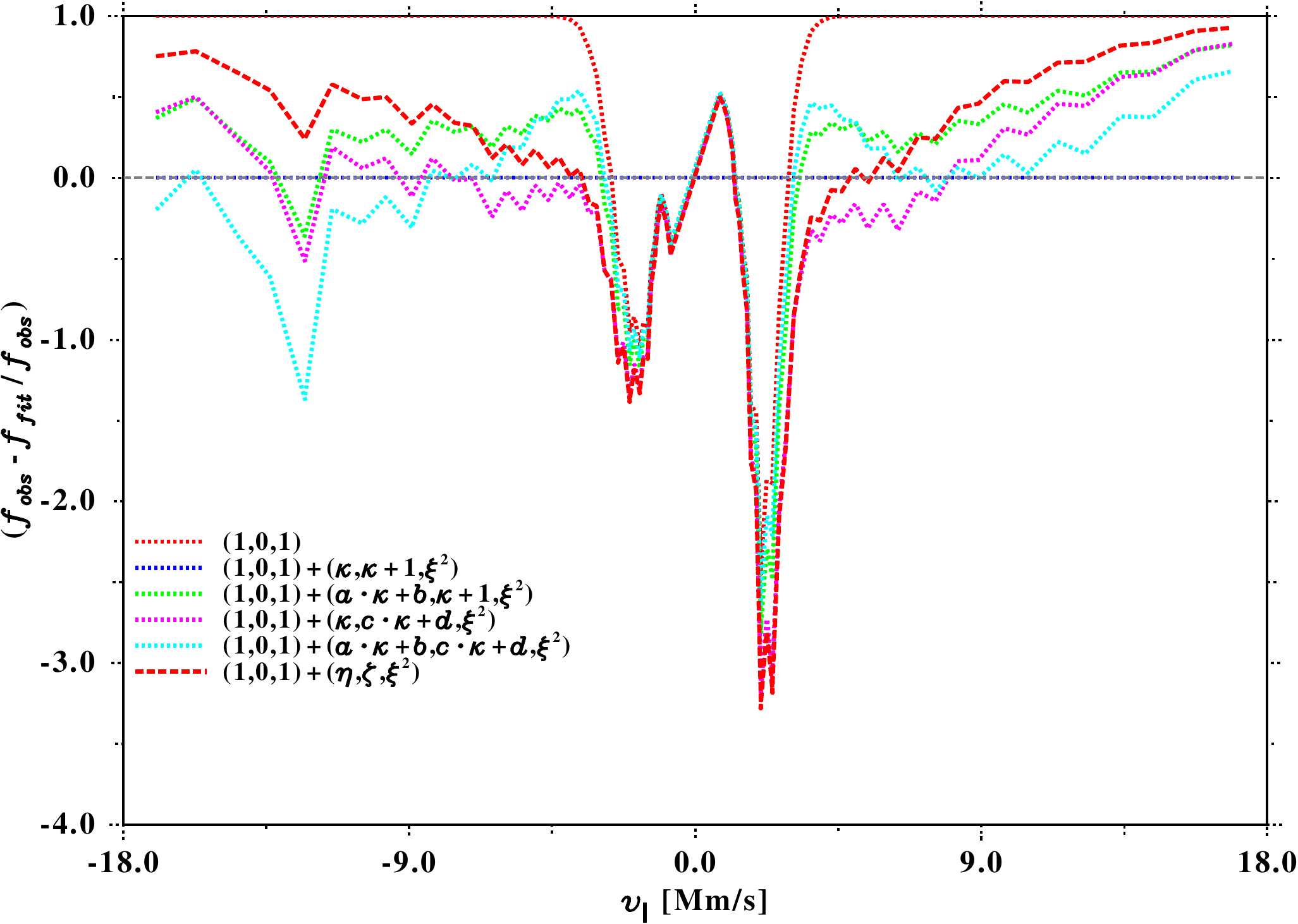}
  \caption{Top panel: The stars mark the Ulysses observations of the parallel
    part of an electron distribution. The recipes fitted are given
    in the legend. Bottom panel: The relative errors of recipes. See main
    text for more information.\label{fig:Ulysses2}}
\end{figure}

We present another data
set from Ulysses (from 19.01.2002 at 08:19:49) where the $\kappa$-value is 
quite low ($\kappa\approx 1.54$). The fit to the data is
shown in the top panel of Fig.~\ref{fig:Ulysses2}, while in the
bottom panel the relative error is presented.
 \begin{table*}
 \scalebox{0.7}{
 \begin{tabular}{l|lll|lllllc|llll}
 Recipe & $n_{M}$ [cm$^{-3}$] & $\Theta_{M}$ [Mm$^{-3}]$& $u_{M}$  [Mm$^{-3}]$& 
 $n_{\kappa}$[cm$^{-3}]$ & $\Theta_{\kappa}$ [Mm$^{-3}]$ & $u_{\kappa}$ [Mm$^{-3
 }]$ & $\eta$& $\zeta$ & $\xi$ & $a$ & $b$ & $c$ & $d$\\ 
 &\multicolumn{3}{c|}{~} & \multicolumn{3}{c}{~}&\multicolumn{2}{c}{$\kappa$} &$
 \xi$ &&&&\\
 \hline
$(1,0,1)$                                                                                            &      0.505 &     1.227 &     0.054 &
 &&&&&&&&\\
$(1,0,1)+(\kappa ,\kappa+1,\xi^2)$                                                                   &      0.507 &     1.234 &     0.054 &     0.100 &     2.159 &     0.167  &\multicolumn{2}{c}{     1.546 } & 0.523E-01  &
 &&&\\
$(1,0,1)+(a\cdot\kappa+b ,\kappa+1,\xi^2)$                                                           &      0.508 &     1.235 &     0.054 &     0.100 &     2.162 &     0.168  &\multicolumn{2}{c}{     1.547 } & 0.524E-01  &     1.031 &     1.120 &           &     0.000 
 \\
$(1,0,1)+(\kappa ,c\cdot\kappa+d,\xi^2)$                                                             &      0.507 &     1.235 &     0.054 &     0.100 &     2.163 &     0.168  &\multicolumn{2}{c}{     1.549 } & 0.523E-01  &           &           &     1.103 &     1.114 
 \\
$(1,0,1)+(a\cdot\kappa+b ,c\cdot\kappa+d,\xi^2)$                                                     &      0.507 &     1.235 &     0.054 &     0.100 &     2.163 &     0.168  &\multicolumn{2}{c}{     1.547 } & 0.523E-01  &     1.720 &     0.820 &     1.255 &     0.570 
 \\
$(1,0,1)+(\eta , \zeta ,\xi^2)$                                                                      &      0.508 &     1.237 &     0.054 &     0.100 &     2.164 &     0.168  &     1.548  &     3.114  & 0.524E-01  &
 &&&\\
 \hline
 $<\bullet>$ &
     0.507 &     1.233 &     0.054 &     0.100 &     2.162 &     0.168 &\multicolumn{2}{c}{     1.547 } & 0.523E-01 &
 &&&\\
 $\sigma(\bullet)$ &
 0.854E-03 & 0.300E-02 & 0.170E-03 & 0.511E-04 & 0.135E-02 &     0.000 &\multicolumn{2}{c}{     0.001 } & 0.277E-04 &
 &&&
 \end{tabular}
 }
 \caption{\label{tab:fit2} Similar to Table.~\ref{tab:fit} for 
 the parallel data from day 19.01.2002 at 08:19:49.
 Here all the recipes with $\xi=0$ and containing the factor $\kappa-\thalf$
 are neglected because of the low $\kappa$ value. See text.}
 \end{table*}
 \begin{table*}
 \scalebox{0.7}{
 \begin{tabular}{l|lll|lll}
 Recipe & $u_{p,M}$ [km/s] & $P_{M} $ [m$^{-1}$s$^{-1}$]& $R_{M}$ [m$^{-1}$s$^{-
 1}$] & $u_{p,\kappa}$[km/s] &  $P_{\kappa} $ [m$^{-1}$s$^{-1}$]& $R_{\kappa}$ [
 m$^{-1}$s$^{-1}$]\\
 \hline
$(1,0,1)$ &      699.4 &     0.507 &    0.0015 \\$(1,0,1)+(\kappa ,\kappa+1,\xi^2)$ &      705.5 &     0.514 &    0.0015 &     336.9 &     0.592 &    0.0028 \\$(1,0,1)+(a\cdot\kappa+b ,\kappa+1,\xi^2)$ &      707.3 &     0.516 &    0.0015 &     307.8 &     0.465 &    0.0028 \\$(1,0,1)+(\kappa ,c\cdot\kappa+d,\xi^2)$ &      706.6 &     0.516 &    0.0015 &     304.1 &     0.493 &    0.0028 \\$(1,0,1)+(a\cdot\kappa+b ,c\cdot\kappa+d,\xi^2)$ &      706.3 &     0.515 &    0.0015 &     297.2 &     0.422 &    0.0028 \\$(1,0,1)+(\eta , \zeta ,\xi^2)$ &      708.2 &     0.517 &    0.0015 &     275.2 &     0.410 &    0.0028 \\
 \hline
 $<\bullet>$ &
     705.5 &     0.514 &    0.0015 &     253.5 &     0.397 &    0.0023 \\
 $\sigma(\bullet)$ &
       2.9 &     0.003 &    0.0000 &     114.8 &     0.187 &    0.0011
 \end{tabular}
 }
 \caption{\label{tab:presfit2} Similar to Table~\ref{tab:presfit2} 
 for the data set from the Ulysses data set at 19.01.2002 at 08:19:49.
 presented. See text for further discussion.
 }
 \end{table*}
Fits are in general much better than the previous one. We did not fit the recipes with 
$\xi=0$ because at such low $\kappa$-values the contribution of superluminal particles to 
the pressure is not negligible \citep{Scherer-etal-2019b}. We have also neglected the 
recipes labeled with $\kappa-1.5$, because of the most probable speed and all the moments 
which become indefinite when $\kappa\to\thalf$. Remarkable is in this case the benefit of 
using RKD-like models, and, nevertheless, the remaining recipes give all similar macroscopic 
quantities, thus one can also use them to fit the data.

\subsubsection{Conclusion from the fits}
From the above fits one can conclude that all these fits of isotropic distribution functions 
(recipes) lead to very similar macroscopic parameters, excepting those which constrain the
existence of these parameters (moments of order $l$) only for sufficiently large values of 
$\kappa$($>(l+1)/2$). The ``bad'' fits to the data points indicate that the choice of an isotropic
distribution function is not sufficient, but anisotropic or more detailed distribution functions,
(e.g., separate halo and strahl) may offer more accurate descriptions. Also, the unrealistically 
high values obtained for the electron number density may have similar explanations. With 2D data
sets, like those used here above, it becomes then possible to extend our present cookbook, by 
including anisotropic recipes, which we intend to write in the future.

\section{Discussion and Conclusion}\label{sec:conclusion}

In this paper we have presented a generalization of the $\kappa$-distributions, called the 
    $\kappa$-cookbook, which unifies the various $\kappa$-distributions known in literature, 
    called recipes. After laying out the required conditions for all distributions, we presented 
    the generalized $\kappa$-distribution (GKD), and discussed its general properties and shape. 
    Special cases with one of the parameters of the GKD being zero were discussed, and some commonly 
    used $\kappa$-distributions were examined with respect to their limits and shape. Subsequently, 
    we presented higher-order moments, i.e., the most probable speed (1. moment) and the pressure 
    (2. moment) for the GKD and commonly used recipes, discussed the Debye length, and, ultimately,
    applied the examined recipes to fit real data measured by Ulysess.

The vanishing of the most probable speeds, 
the pressure (except for the recipe $(\kappa-\thalf, \kappa+1, 0)$) and Debye lengths
for the recipes with $\eta=\kappa \to \thalf$
contradicts the physical interpretation that the distributions for
rarefied gases ``far'' from equilibrium do have high most probable
speeds or pressure like the SKD and RKD. 

Even worse is the recipe $(\kappa-\thalf, \kappa+1, 0)$,
because here we have a constant pressure, but a vanishing most probable
speed for $\kappa \to \thalf$.
That means that the recipe $(\kappa-\thalf, \kappa+1, 0)$ has a finite pressure value, but no average
speed.  That behavior can be expected in a crystal at low temperatures,
but not in a (rarefied) gas or plasma.

The recipe $(\kappa-\thalf, \kappa+1, 0)$ and SKD have the same problem concerning higher-order moments:
the critical $\kappa$-value must increase with the order to get a finite
moment, e.g.,\ for the third moment $\kappa>2$ is required.
The higher-order moments for the recipes with $\eta=\kappa - \thalf$
vanish when $\kappa\to\thalf$, and thus these recipes with values close
to that have problems, especially when $\xi=0$, because then the
contribution from superluminal particles becomes too large. 

Let us assume that we have two distributions, one Maxwellian and another one constant.
For the Maxwellian all moments exist, and for the constant
distribution they are divergent or indefinite (do not exist). Thus, a general distribution
with high-energy tails but Maxwellian core should have higher-order moments than the
Maxwellian, because for lower $\kappa$-values
the form part approaches a constant, but the distribution function becomes indefinite due to the normalization. With increasing $\kappa$
the tails becomes steeper and the moments decrease and reach that of a
Maxwellian.


It becomes thus clear that the RKD provides a practical alternative recipe for defining macroscopic moments of the 
    observed distributions without any constraint for the power-law exponent $\kappa$. Our final result shows that all  
    discussed recipes lead to similar macroscopic velocity moments, providing quantities like number density, core and
    and drift speeds. Therefore, fitting the observed data with one of the above kappa recipes will not change the
    macroscopic behavior of the plasma and thus we have not to care about large scale fluid models (of the heliosphere), because they are all similar for the discussed distribution functions.
    Nevertheless, the microscopic (kinetic) behavior is affected by the choice of a recipe 
    \citep{Yoon-etal-2019, Husidic-etal-2020}. Now when defining a new recipe, say, to better fit the data, it is easy
    with the above derived formulas to calculate the velocity moments, Debye length etc. and compare them with the
    standard recipes for the SKD or RKD.
\section{Data availability statement}

The datasets were derived from sources in the public domain: \verb+https://omniweb.gsfc.nasa.gov/+
\bibliographystyle{mnras}
\bibliography{test}

\appendix
\section{Some useful properties of the Kummer-$U$ or Tricomi function}
\subsection{Useful properties}
The $n$th
derivative with respect to $x$
of $U(a,b,x)$ is \citep[][Eq.~(13.4.22)]{Abramowitz-Stegun-1972}

\begin{align}
  \frac{d^{n}}{d\xi^{n}} U(a,b,\eta\xi) = (-\eta)^{n} \frac{\GA{a+n}}{\GA{a}}U(a+n,b+n,\eta\xi)\,,
\end{align}
and thus for the even moments we find for the derivative with respect
to $\xi$ (from Eq.~(\ref{gkdnorm}))
\begin{align}\nonumber
  U&\left(\thalf,\frac{5}{2}-\zeta,\eta\xi\right) \tilde{M}_{2n}
  = \eta^{n}
          U\left(\frac{3}{2}+n,\frac{5}{2}+n-\zeta,\xi\eta\right)\\\nonumber
  &\qquad= (-1)^{n}\frac{\GA{\thalf}}{\GA{\thalf+n}}\frac{d^{n}}{d\xi^{n}}
    U\left(\thalf,\frac{5}{2}-\zeta,\eta\xi\right)\\
  &\qquad= -(1)^{n} \frac{\sqrt{\pi}}{2 \GA{\thalf+n}}
    \frac{d^{n}}{d\xi^{n}} U\left(\thalf,\frac{5}{2}-\zeta,\eta\xi\right)\,,
\end{align}
and for the even moments
\begin{align}\nonumber
  &U\left(\thalf,\frac{5}{2}-\zeta,\eta\xi\right) \tilde{M}_{2n+1}
  = \eta^{\half+n}U\left(2+n,3+n-\zeta,\xi\eta\right)\\\nonumber
  &\qquad = (-1)^{n}\eta^{\half} \frac{\GA{2}}{\GA{2+n}}\frac{d^{n}}{d\xi^{n}}
    U\left(2,3-\zeta,\eta\xi\right)\\
  &\qquad = (-1)^{n}\eta^{\half}\frac{1}{\GA{2+n}}
   \frac{d^{n}}{d\xi^{n}} U\left(2,3-\zeta,\eta\xi\right)\,.
\end{align}
Especially for the pressure we find:
\begin{align}
  \tilde{M}_{2} = -\frac{2}{3} \,\frac{d}{d\xi}\ln\left[U\left(\thalf,\frac{5}{2}-\zeta,\eta\xi\right)\right]\,.
\end{align}

To evaluate Eq.~(\ref{eq:16}) we go to the integral
representation of the \KUT, see \citet{Abramowitz-Stegun-1972} Eq.~(13.2.5):
\begin{align}\label{eq:aa}\nonumber
    \GA{\frac{n+3}{2}} &U\left(\frac{n+3}{2},\frac{n+5}{2},1\right) =
    \int\limits^{\infty}_{0} e^{-t} t^{\frac{3+n}{2}-1} 
      (1+t)^{\frac{5+n}{2}-\frac{3+n}{2}-1} \md t  \\
                      &=\int\limits^{\infty}_{0} e^{-t} t^{\frac{3+n}{2}-1}\md t 
    =\GA{\frac{n+3}{2}}
\end{align}
and thus $U\left(\frac{n+3}{2},\frac{n+5}{2},1\right)=1$.

\subsection{Limits\label{limits}}

The limits for low values of the third argument in the Kummer-$U$ function, can be found in
\citet[][Eq.s~(13.5.10) to (13.5.12)]{Abramowitz-Stegun-1972}. 
In Table~\ref{tab:U_cases} the limits are
more involved when we use $\eta=\eta(\kappa)$ and
$\zeta=\zeta(\kappa)$ and want to study the case when $\kappa\to
0$. The recipes have to be checked individually. 

Thus, for the limits when $\xi\to 0$ we can combine case~1 and
case~3 to: 
\begin{align}\label{eq:ulim}
U\left(\frac{3+n}{2},\frac{5+n}{2}-\zeta,0\right) &=
         \frac{\GA{\zeta-\frac{3+n}{2}}}{\GA{\zeta}}  \qquad
                                         \mathrm{if\ \ }\zeta > \frac{3+n}{2} \,,
\end{align}
and for the moments $\tilde{M}^{n}$ we find:
\begin{align}\nonumber
   \lim\limits_{\xi\to 0}\eta^{\frac{n}{2}} &\frac{
          U\left(\frac{n+3}{2},\frac{n+5}{2}-\zeta,\xi\eta\right)}
  { U\left(\frac{3}{2},\frac{5}{2}-\zeta,\xi\eta\right)} \\
  &=
  \eta^{\frac{n}{2}}
  \frac{\GA{\zeta-\frac{3+n}{2}}}{\GA{\zeta-\thalf}} \qquad \mathrm{if\ \ }
  \zeta > \frac{3+n}{2}\,.
\end{align}

For the recipes ($\kappa,\kappa+1,\xi=\const$) we have to use case~5 
for the normalization, case~6
for the most probable speed and case~7 for the pressure:
\begin{align}\nonumber
  \lim\limits_{\kappa\to 0} \tilde{u}_{p} &=
            \lim\limits_{\kappa\to 0} \sqrt{\kappa}
                                            \Uka[1][0](\kappa,\kappa+1,\xi)
  \\\label{good}
  &=  \lim\limits_{\kappa\to 0}
            \sqrt{\kappa}\, \frac{\frac{1}{\kappa\xi}+
         \ln(\kappa\xi)}
           {\frac{\GA{\frac{1}{2}-\kappa}}{\GA{\frac{3}{2}}(\kappa\xi)^{\frac{1}{2}-\kappa}}
  +\frac{\GA{-\frac{1}{2}+\kappa}}{\GA{\kappa+1}} } =
            \frac{1}{2\sqrt{\xi}} \\\nonumber
  \lim\limits_{\kappa\to 0} \tilde{P}& = \lim\limits_{\kappa\to 0} \kappa \Uka[2][0](\kappa,\kappa+1,\xi)
  \\\label{limp}
  &= \lim\limits_{\kappa\to 0}
           \kappa \frac{\frac{\frac{1}{2}-\kappa-(\kappa+1)\xi\kappa}
           {\GA{\frac{5}{2}}(\kappa\xi)^{\frac{3}{2}-\kappa}}
                                      \GA{\half-\kappa}}
            {\frac{\GA{\frac{1}{2}-\kappa}}{\GA{\frac{3}{2}}(\kappa\xi)^{\frac{1}{2}-\kappa}}
  +\frac{\GA{-\frac{1}{2}+\kappa}}{\GA{\kappa+1}} } = \frac{1}{3\xi}\,,
\end{align}
while for the recipes ($\kappa-\thalf,\kappa+1,\xi=\const$) we have to
choose case~2 for the normalization and cases~3 and~4 for the most
probable speed and pressure, when $\kappa\to\thalf$.  
\begin{align}\nonumber
    \lim\limits_{\kappa\to \thalf} \tilde{u}_{p} &=
            \lim\limits_{\kappa\to \thalf} \sqrt{\kappa-\half}
          \Uka[1][0](\kappa-\thalf,\kappa,\xi)\\\nonumber
            & = \lim\limits_{\kappa\to \thalf} \sqrt{\kappa-\thalf}                      \frac{
          \frac{\GA{\kappa}{\GA{\kappa+1}}} + \frac{\GA{-\kappa}}{\GA{3}}([\kappa-\thalf]\xi)^{-\kappa}}
            {\frac{1+\thalf                         (\kappa-\thalf)\xi\log([\kappa-\thalf]\xi)}{\GA{3}}} =0\\
   \lim\limits_{\kappa\to \thalf} \tilde{P}& = \lim\limits_{\kappa\to \thalf} \kappa \Uka[2][0](\kappa-\thalf,\kappa+1,\xi)\\
          &= \lim\limits_{\kappa\to \thalf}(\kappa-\thalf)\frac{
          \frac{2\gamma+\psi\left(\frac{5}{2}\right)+
           \ln([\kappa-\thalf]\xi)}{-\GA{\frac{5}{2}}}}
           {\frac{1+\thalf
             (\kappa-\thalf)\xi\log([\kappa-\thalf]\xi)}{\GA{3}}}         = 0 \,.
\end{align}

It can easily be seen that for the recipe
($\kappa-\thalf,\kappa+1,\frac{\kappa\xi}{\kappa-\thalf}$)
the double fraction becomes constant when $\kappa\to \thalf$,
and the factor in front of it goes to zero (similar for higher-order
moments or different recipes).

For the recipe $\kappa-\thalf,\kappa+1,0$   we can combine some of the factors:
\begin{align*}
  \lim\limits_{\kappa\to \thalf}& f(\kappa-\thalf, \kappa+1, 0)\\
                =& \lim\limits_{\kappa\to \thalf}  \frac{\kappa\Gamma(\kappa)}{\Gamma\left(\kappa-\frac{1}{2}\right)\left(\kappa-\frac{3}{2}
    \right)^{\frac{3}{2}}\sqrt{\pi^{3}}\Theta^{3}} 
  \left(1 +
  \frac{v^{2}}{\left(\kappa-\frac{3}{2}\right)\Theta^{2}}\right)^{-\kappa-1}\\
  =&\thalf \frac{\GA{\thalf}}{\GA{1}\sqrt{\pi^{3}}\Theta^{3}}
     \lim\limits_{\kappa\to \thalf}
     \frac{\left(\kappa-\thalf\right)^{\kappa+1}}{\left(\kappa-\thalf\right)^{\thalf}}
     \left(\kappa-\frac{3}{2} +
     \frac{v^{2}}{\Theta^{2}}\right)^{-\kappa-1}\\
      =& \thalf\frac{\GA{\thalf}}{\GA{1}\sqrt{\pi^{3}}\Theta^{3}}
         \left(\frac{v^{2}}{\Theta^{2}}\right)^{-\frac{5}{2}}
        \lim\limits_{\kappa\to \thalf}
     \left(\kappa-\thalf\right)
     = 0
\end{align*}
with $\lim\limits_{\kappa\to 0}\kappa^{\kappa} = 1$.

\subsection{The limits $\kappa\to\infty$}
These limits are more complicated and it is the best to calculate the
limiting distribution function and from that the limiting moments.

To calculate the upper limits of the normalization factor for the
recipe ($\kappa,\kappa+1,\xi^{2}$), the RKD, we use the original integral:
\begin{align*}
\lim\limits_{\kappa\to\infty}&\kappa^{\frac{3}{2}} U\left(\thalf,
                               \thalf-\kappa,\kappa\xi^{2}\right) \\
  &=
    \lim\limits_{\kappa\to\infty}
   \frac{1}{2\GA{\thalf}}\int\limits^{\infty}_{0}
   e^{-\frac{\xi^{2}v^{2}}{\Theta^{2}}}\left(1+\frac{v^{2}}{\kappa\theta^{2}}\right)^{-\kappa-1}v^{2}
   \md v\\
   &=  \frac{1}{2\GA{\thalf}}\int\limits^{\infty}_{0}
           e^{-\frac{1+\xi^{2}v^{2}}{\Theta^{2}}} v^{2}\md v              
=  \frac{\sqrt{\pi}\Theta^{3}}{4\left(1+\xi^{2}\right)^{\thalf}}\,,
\end{align*}
which has to be multiplied by $4\pi$ because of the spherical volume
element. In an analogous  way one gets the upper limits of the other
distribution functions.
\begin{table}
\begin{tabular}{l|K|K}\label{tab:limit}
  Case 1:
  & b<0
  & \frac{(b+a \eta\xi)\GA{-b}}{-\GA{\zeta}} \\
  Case 2:
  & b = 0
  & \frac{1+ a \eta\xi \ln(\eta\xi)}{\GA{a+1}} \\
  Case 3:
  & 0 <b <1
  & \frac{\GA{1-b}}{\GA{\zeta}}+\frac{\GA{b-1}(\eta\xi)^{1-b}}{\GA{a}}\\
  Case 4:
  & b= 1
  & \frac{2\gamma + \psi(a)+\ln{\eta\xi}}{-\GA{a}} \\         
  Case 5:
  & 1<b<2
  & \frac{\GA{b-1}}{\GA{a}(\eta\xi)^{b-1}}+\frac{\GA{1-b}}{\GA{\zeta}}\\
  Case 6:
  & b = 2
  & \frac{1}{\GA{a}\eta\xi} + \frac{\ln{\eta\xi}}{\GA{a-1}}\\
  Case 7:
  & b>2
  & \frac{(b-2-\zeta \eta\xi)\GA{b-2}}{\GA{a}(\eta\xi)^{b-1}}  
\end{tabular}
\caption{For $U(a,b,x)$ with $a=\frac{3+n}{2}$ and
  $b=\frac{5+n}{2}-\zeta$, where $\gamma\approx 0.5772$ is the Euler constant, and
$\psi$ is the digamma function.}\label{tab:U_cases}
\end{table}

\subsection{The case $\eta \to 0$}\label{sec:eta0}
From the above we find
\begin{align}\nonumber
  \lim\limits_{\eta\to 0} \tilde{f} =&
  e^{-\xi(\kappa)\frac{v^{2}}{\Theta^{2}}}
  \lim\limits_{\eta\to 0}
\frac{1}{\eta^{\thalf}U(\thalf,\frac{5}{2}-\zeta,\eta\xi)}
  \left(1+\frac{v^{2}}{\eta\Theta^{2}}\right)^{-\zeta}\\\nonumber
  =& e^{-\xi(\kappa)\frac{v^{2}}{\Theta^{2}}}
  \lim\limits_{\eta\to 0}
\frac{\eta^{\zeta-\thalf}}{U(\thalf,\frac{5}{2}-\zeta,\eta\xi)}
  \left(\eta+\frac{v^{2}}{\Theta^{2}}\right)^{-\zeta}\\\nonumber
 =&  e^{-\xi(\kappa)\frac{v^{2}}{\Theta^{2}}}
    \lim\limits_{\eta\to 0} \eta^{\zeta-\thalf}
    \left(\eta+\frac{v^{2}}{\Theta^{2}}\right)^{-\zeta}\\\nonumber
   &\cdot    \left\lbrace
    \begin{array}{lr}
      \frac{\sqrt{\pi}\eta^{\thalf}\xi^{\thalf}\GA{\zeta}}
   {2\eta^{\zeta}\GA{\thalf-\zeta}\GA{\zeta}+\sqrt{\pi}\eta^{\thalf}\xi^{\thalf}\GA{\zeta-\thalf}}
       & \hspace*{1cm}\zeta > \thalf\\
       \frac{3\sqrt{\pi}}{4[1+\eta\xi*\ln(\eta\xi)]}
       & \zeta = \frac{5}{2}\\
       \frac{2\GA{\zeta}}{(2\zeta-3\eta\xi-5)\GA{\zeta-\frac{5}{2}}}
       & \zeta > \frac{5}{2} 
    \end{array}
\right.\\
  &=0 \,.
\end{align}
The first line goes to zero, because the denominator stays finite,
while the numerator goes to zero. The same holds true for the second
line ($\lim\limits_{x\to 0}x\ln(x)=0$). The third row goes to zero
because $\zeta>5$ and the denominator is positive. For $\zeta\le
\thalf$ the distribution function goes to infinity.

The moments are all proportional to $\eta$ and go in general to zero,
with the exceptional case discussed in subsection~\ref{sec:zeta}, where
one moment can be chosen to be finite:
\begin{align}\nonumber
  \lim\limits_{\eta\to 0} & \tilde{M}_{n}= \lim\limits_{\eta\to 0} \eta^{\frac{n}{2}}
  \frac{U\left(\frac{n+3}{2},\frac{n+5}{2}-\zeta,\eta\xi\right)}
 {U\left(\frac{3}{2},\frac{5}{2}-\zeta,\eta\xi\right)}
                 =  \lim\limits_{\eta\to 0} \eta^{\frac{n}{2}}
    \\\nonumber&\cdot\left\lbrace
    \begin{array}{lr}
       \frac{\sqrt{\pi}}{\GA{\frac{n+3}{2}}}\,
        \frac{\GA{\frac{3+n}{2}-\zeta}(\eta\zeta)^{\zeta-\frac{3+n}{2}}
        \GA{\zeta}+\GA{\zeta-\frac{3+n}{2}}\GA{\frac{n+3}{2}}}
        {2\GA{\thalf-\zeta}\left(\eta\zeta\right)^{\zeta-\thalf}
        \GA{\zeta}+ \sqrt{\pi}\GA{\zeta-\thalf}}
      & \hspace*{0.25cm}\zeta > \thalf\\
  \frac{\sqrt{\pi}\GA{\zeta}}{2\GA{\frac{n+5}{2}}}\,
  \frac{2+\eta\xi(n+3)\ln(\eta\xi)}
  {2\GA{\thalf-\zeta}\left(\eta\zeta\right)^{\zeta-\thalf}\GA{\zeta}+
  \sqrt{\pi}\GA{\zeta-\thalf}}
                                        &\zeta = \frac{5}{2}\\
  -\frac{\GA{\zeta-\frac{5+n}{2}}}{\GA{\zeta}}\left(\frac{3+n}{2}\eta\xi+\frac{5+n}{2}-\zeta\right)
   &\zeta > \frac{5}{2}\\
    \end{array}
  \right.\\\nonumber
  &=0 \,.
\end{align}
The fractions in the first and second line become constant when
$\eta\to 0$, because the numerator and denominator have both a term
independent of $\eta$, and thus the limit goes to zero. The factor in
the third line has no dependence in the denominator on $\eta$, and thus
vanishes also.

\section{The integrals}\label{integrals}
We calculate the moments $M_{n}=N_{G}^{-1}M_{n}^{\prime}$ of the GKD by:
\begin{align}\nonumber
  M_{n}^{\prime} &= 4\pi\int\limits_{0}^{\infty}\left(1+\frac{v^{2}}{\eta(\kappa)\Theta^{2}}\right)^{-\zeta(\kappa)}e^{-\xi(\kappa)\frac{v^{2}}{\Theta^{2}}}
                   v^{2+n} \md v\\\nonumber
        &= 2\pi\eta^{\frac{3+n}{2}}\Theta^{3+n}
          \int\limits_{0}^{\infty}\left(1+x\right)^{-\zeta}
          e^{-\xi \eta x}x^{\frac{1+n}{2}} \md x\\
  &=  2\pi \eta^{\frac{3+n}{2}}\Theta^{3+n}\GA{\frac{n+3}{2}} U\left(\frac{n+3}{2},\frac{n+5}{2}-\zeta,\xi\eta\right)
\end{align}
If $\xi=0, \zeta=\kappa+1, \eta=\kappa$ we obtain the SKD, etc., where
the limiting cases (see Appendix~\ref{limits}) have to be taken into account whenever the
argument $\xi\eta$ of the Kummer-$U$ function approaches zero.

The normalization of the GKD is given by (including the factor $4\pi$ from
the spherical volume element)
\begin{align}
  N_{G}^{-1} = \tilde{M}_{0}= \eta^{\thalf}\Theta^{3}\sqrt{\pi^{3}} U\left(\thalf,\frac{5}{2}-\zeta,\xi\eta\right)\,.
\end{align}

\subsection{The entropy for the GKD}\label{entropy}
The normalized entropy $S = S' /k_{B}$ ($k_{B}$ is the Boltzmann
constant and $S'$ the entropy) is given by Boltzmann's H-theorem by (with $4 \pi$
from the spherical
volume element) and neglecting the Gibbs correction (and the ``-1'' part)
\begin{align}\label{eq:ent0}
  S =& -\int\int f \ln f\, \md^{3}v\,\md^{3}x\, \\\nonumber
\end{align}
and from the above
  \begin{align}\nonumber
 S =& - \frac{4 \pi}
{\eta^{\thalf}\Theta^{3}\sqrt{\pi^{3}} U\left(\thalf,\frac{5}{2}-\zeta,\xi\eta\right)}
\int n_{0} \int\limits_{0}^{\infty}\left(1+\frac{v^{2}}{\eta(\kappa)\Theta^{2}}\right)^{-\zeta(\kappa)}\\\nonumber
    &\cdot e^{-\xi(\kappa)\frac{v^{2}}{\Theta^{2}}}\ln\left(
\frac{n_{0} \left(1+\frac{v^{2}}{\eta(\kappa)\Theta^{2}}\right)^{-\zeta(\kappa)}e^{-\xi(\kappa)\frac{v^{2}}{\Theta^{2}}}}
{\eta^{\thalf}\Theta^{3}\sqrt{\pi^{3}} U\left(\thalf,\frac{5}{2}-\zeta,\xi\eta\right)}
     \right)
    v^{2}\md v\, \md^{3} x \\\nonumber
  =& - 4\pi \int \int\limits_{0}^{\infty} f
     \ln\left(
     \frac{ n_{0} \left(1+\frac{v^{2}}{\eta(\kappa)\Theta^{2}}\right)^{-\zeta(\kappa)}e^{-\xi(\kappa)\frac{v^{2}}{\Theta^{2}}}
     }
{\eta^{\thalf}\Theta^{3}\sqrt{\pi^{3}} U\left(\thalf,\frac{5}{2}-\zeta,\xi\eta\right)}
     \right)
     v^{2}\md v\, \md^{3} x \\\nonumber
  =& - 4\pi \int \int\limits_{0}^{\infty} f \ln\left(\frac{n_{0}}
     {\eta^{\thalf}\Theta^{3}\sqrt{\pi^{3}}
     U\left(\thalf,\frac{5}{2}-\zeta,\xi\eta\right)}\right)
     v^{2}\md v\, \md^{3} x  \\\label{eq:ent}
     &  - 4\pi \int \int\limits_{0}^{\infty} f \ln\left(
       \left(1+\frac{v^{2}}{\eta(\kappa)\Theta^{2}}\right)^{-\zeta(\kappa)}e^{-\xi(\kappa)\frac{v^{2}}{\Theta^{2}}}
       \right)
     v^{2}\md v\, \md^{3} x \,.
\end{align}
In the first integral above the factor inside the $\ln$ is independent of
$v$. We split the first integral further by evaluating the
logarithm. Then one integral is 
\begin{align}
  \int n_{0}\ln n_{0} \md^{3} x = N\ln \frac{N}{V} + N \int \frac{d \, \ln n_{0}}{dx}\, \md^{3} x \approx N \ln n_{0} + \const\,,
\end{align}
which gives approximately the total number of particles $N$.

All other integrals are of the type $\int n_{0}\int G(v)\md
v\,\md^{3}x$, which gives the total number of particles $N$  when
integrating over the volume, and the
remaining parts are integrals with respect to $v$.
With $\tilde{S}= S / N -\const $ and
with $\tilde{f}= f / n_{0}$ we find then:
\begin{align}\nonumber
  \tilde{S} =& - 4\pi \left(
               \ln n_{0} - 3 \ln \Theta -
                \ln\left[\eta^{\thalf}\sqrt{\pi^{3}}
      U\left(\thalf,\frac{5}{2}-\zeta,\xi\eta\right) \right]
               \right)\\\nonumber
& - 4\pi \left(
  \underbrace{-\zeta \int\limits_{0}^{\infty} \tilde{f}
  \ln\left[1+\frac{v^{2}}{\eta\Theta^{2}}\right] v^{2}\md v}_{\mathrm{The\ log-term}}
-\frac{\xi}{\Theta^{2}}\int\limits_{0}^{\infty} \tilde{f}
 v^{4}\md v
  \right)\,.
\end{align}
The last expression is the pressure divided by $n_{0}$ and we insert Eq.~(\ref{gpres})
(without $n_{0}$). 
To continue we replace in the log-term 1 by $\epsilon$, differentiate
with respect to $\epsilon$, expand into a Taylor series around
$\epsilon=1$,  integrate, and set $\epsilon = 1$:
\begin{align}\nonumber
& \left. -\zeta \int\limits_{\epsilon}\frac{\partial}{\partial \epsilon}\int\limits_{0}^{\infty} \tilde{f}
  \ln\left[\epsilon+\frac{v^{2}}{\eta\Theta^{2}}\right] v^{2}\md v\, \md
  \epsilon\right|_{\epsilon=1}\\\nonumber
  &= \left. -\zeta \int\limits_{\epsilon}\int\limits_{0}^{\infty}
    \tilde{f}\frac{1}{\epsilon+\frac{v^{2}}{\eta\Theta^{2}}} v^{2}\md
    v\, \md \epsilon\right|_{\epsilon=1}\\\nonumber
   &=\left.  -\zeta \int\limits_{\epsilon} \int\limits_{0}^{\infty}
    \tilde{f}\sum\limits_{l=0}(-1)^{l}\frac{(\epsilon-1)^{l}}{(1+\frac{v^{2}}{\eta\Theta^{2}})^{l+1}}v^{2}\md
     v\, \md \epsilon\right|_{\epsilon=1}\\\nonumber
  &=\left.  -\zeta \int\limits_{0}^{\infty}\tilde{f}\sum\limits_{l=0}(-1)^{l}\frac{(\epsilon-1)^{l+1}+(-1)^{l}}{(l+1)(1+\frac{v^{2}}{\eta\Theta^{2}})^{l+1}}v^{2}\md
    v \right|_{\epsilon=1}\\\nonumber
    &=  -\zeta
      \int\limits_{0}^{\infty}\tilde{f}\sum\limits_{l=0}\frac{1}{(l+1)(1+\frac{v^{2}}{\eta\Theta^{2}})^{l+1}}v^{2}\md v \\\nonumber
&=-\frac{\zeta}{\sqrt{\pi^{3}}\eta^{\thalf}\Theta^{3}U\left(\thalf,\frac{5}{2}-\zeta,\eta\xi\right)}\\\nonumber
  &\hspace{1cm}\cdot\int\limits_{0}^{\infty} \sum\limits_{l=0}\frac{1}{l+1}
  \left(1+\frac{v^{2}}{\eta\Theta^{2}}\right)^{-\zeta-l-1}e^{-\xi\frac{v^{2}}{\Theta^{2}}}
  v^{2}\md v \\
&=-\frac{\zeta}{2\pi} \sum\limits_{l=0}\frac{1}{l+1}\,
  \frac{U\left(\thalf,\half-\zeta-l,\eta\xi\right)}
  {U\left(\thalf,\frac{5}{2}-\zeta,\eta\xi\right)}\,,
\end{align}
from which follows

\begin{align}\nonumber
   \tilde{S} =& - 4\pi \left(
               \ln n_{0} - 3 \ln \Theta -
                \ln\left[\eta^{\thalf}\sqrt{\pi^{3}}
      U\left(\thalf,\frac{5}{2}-\zeta,\xi\eta\right) \right]
                \right)\\\nonumber
  &- 4\pi \left(
  -\frac{\zeta}{2\pi} \sum\limits_{l=0}\frac{1}{l+1}\,
  \frac{U\left(\thalf,\half-\zeta-l,\eta\xi\right)}
  {U\left(\thalf,\frac{5}{2}-\zeta,\eta\xi\right)}\right.\\\label{eq:entgkd}
  &\hspace{1cm}\left.-\frac{2\xi\eta}{3}\,
  \frac{U\left(\frac{5}{2},\frac{7}{2}-\zeta,\xi\eta\right)}
       {U\left(\frac{3}{2},\frac{5}{2}-\zeta,\xi\eta\right)}
    \right)\,.
\end{align}

 For the Maxwellian $\zeta=0, \xi=\eta=1$ we find from
 Eq.~(\ref{eq:entgkd}) and the corresponding limits:
 \begin{align}\label{eq:entmax}
   \tilde{S} =& - 4\pi \ln n_{0} + 12\pi \ln\Theta +\thalf \ln\pi
               +\frac{8\pi}{3}\\\nonumber
  \mathrm{or\ \ \ }\frac{\tilde{S}}{4\pi} =& 3 \ln\Theta - \ln n_{0} + \const \,.
 \end{align}
Replacing $\Theta$ by the thermal speed we get the Maxwellian entropy
(up to a constant).

Even if the Maxwellian part disappears ($\xi=0$), we get a solution
because of ($\zeta>\frac{5}{2}$), and thus the  
the second moment (pressure) exists (last term  of Eq.~\ref{eq:entgkd}).

\section{The integrals for the Debye length}\label{Debye}
The \DL is defined via the Poisson equation
\begin{align}\label{poisson}
  \Delta \Phi & = -\frac{1}{\epsilon_{0}} \left(q_{T}\delta(r) +
                \sum\limits_{s}q_{s}n_{s}(r)\right)  \,,
\end{align}
where $\epsilon_{0}$ is the vacuum permittivity,
$q_{T}$ the charge of a
test particle, $n_{s}$ the disturbed number density of particle
species $s$ (0th moment of the corresponding 
disturbed distribution function) and $q_{s}$ their charge. Because of
charge neutrality $\sum\limits_{s}q_{s}n_{s} = q_{e} n_{0,e} +
q_{i}n_{0,i}=0$ (we only discuss electrons and positively charged ions
(protons), which is stated in the last step).
We rearrange Eq.~(\ref{poisson}) to
\begin{align}
  \Delta \Phi - \frac{1}{\epsilon_{0}}\sum\limits_{s}q_{s}n_{s}(r) = -\frac{q_{T}}{\epsilon_{0}}\delta(r)
\end{align}
and find (for each species)
\begin{align}\label{def}
  \Delta \Phi + \frac{1}{\epsilon_{0}}\sum\limits_{s}q_{s}n_{s}(r)
  \approx
\Delta \Phi + \frac{1}{\epsilon_{0}}\sum\limits_{s}n'_{s}\frac{q^{2}_{s}\Phi}{m_{s}\Theta_{s}^{2}}
  = -\frac{q_{T}}{\epsilon_{0}}\delta(r)\,,
\end{align}
where $m_{s}$ and $\Theta_{s}$ are the mass and thermal core speed for
each species, respectively.
The \DL $\Lambda$ is then defined as
\begin{align}\label{eq:dl}
  \Lambda^{-2} = \Lambda_{e}^{-2} + \Lambda_{i}^{-2} =
  \frac{q_{e}^{2}n'_{e}}{\epsilon_{0}m_{e}\Theta_{e}^{2}} +
  \frac{q_{i}^{2}n'_{i}}{\epsilon_{0}m_{i}\Theta_{i}^{2}}
\end{align}
(for electrons and ions only). The 0th-order moment (number density)
of the perturbed distribution functions is calculated in
Appendix~\ref{Debye}. These perturbed number densities $n_{p}$
are developed into a Taylor series up to the first order in
$\chi=\frac{q\Phi}{m\Theta^{2}}$
(neglecting the index $s$) and yield, in general:
\begin{align}\label{pert}
  n_{s} = n_{0}\left(1 + \frac{q\Phi}{m \Theta^{2}}n'_{s}\right)\,,
\end{align}
where $n'_{s}$
is a factor different for each distribution function. By inserting Eq.~(\ref{pert}) in
Eq.~(\ref{def}) it becomes evident that the first part of Eq.~(\ref{pert})
cancels for ions and electrons species because of the charge neutrality.

To calculate the \DL, we replace the velocity by the kinetic
energy $E_{kin}$
and add as a perturbation the potential energy $V$
(or ``chemical'' potential as in \citet{Treumann-etal-2004} and
\citet{Fahr-Heyl-2016}) of the test particle:
\begin{align}
  v^{2} \to \frac{E_{kin}}{2 m_{s}} \to \frac{E_{kin}+ V}{2 m_{s}}\,,
\end{align}
where $V=q_{s}\Phi$
is the electric potential of a test charge, and
$\pm v_{\Phi}^{2}=q_{s}\Phi/(2 m_{s})$
is the corresponding speed depending on the charge sign. We always
assume that $v_{\Phi}^{2}/\Theta^{2}$
is small.  In contrast to the previous attempts in literature, we replace the
Maxwellian temperature by the the core speed $\Theta$.
The integrals (zeroth moments) corresponding to  the perturbed
distribution functions from Table~\ref{tab:3} and their limits found
in Table~\ref{tab:3a} are discussed in the Appendix~\ref{Debye}. In the \DL the
corresponding $\kappa$-pressure (or temperature) appears as a correction, 
except for the SKD and recipe $(\kappa-\thalf, \kappa+1, 0)$.

We do not follow the Debye-H\"uckel theory, because we want to include
collisionless plasmas, thus we will use the approach by \citet{Krall-Trivelpiece-1973}.

\subsection{The Debye length after Krall \& Trivelpiece}\label{subsec:krall_trivelpiece}

In \citet{Krall-Trivelpiece-1973}, chapter 11.1  a shielding length is
calculated on the basis of a Vlasov-plasma, but for arbitrary
distributions (not too far away from equilibrium), splitting the
distributions functions in
\begin{align}
  f_{s}= f_{0,s} + f_{1,s}\,,
\end{align}
where $f_{0,s}$ is the unperturbed distribution and $f_{1,s}$ the perturbation of
the distribution function for species $s$. In the following we drop
the index $s$.

Following \citep{Krall-Trivelpiece-1973} we have in lowest order that  a
moving test charge follows a straight line
\begin{align}
\vec{x}' =  \vec{x}'_{0}-\vec{v}'t\,,
\end{align}
where $\vec{v}'$ is a uniform velocity and $\vec{x}_{0}'$ is the
location of the test charge at $t=0$. Then the potential can be
written as:
\begin{align}\label{p1}
  \nabla^{2}\Phi = \frac{1}{\epsilon}q \delta(\vec{x}-\vec{x}'_{0}-\vec{v}'t) +
  \frac{1}{\epsilon} n
\end{align}
with
\begin{align}
  n = \sum\limits_{s}q_{s}\int f_{s}\md^{3}v \,.
\end{align}
Note: In our notation the number density $n_{s}$
is part of the distribution function $f_{s}$. The plasma distribution
function $f_{s}$ satisfies the Vlasov equation:
\begin{align}
  \dt{f_{s}} + (\vec{v}\cdot\vec{\nabla})f_{s} -
  \frac{q_{s}}{m_{s}}(\vec{\nabla}\Phi)\cdot \vec{\nabla}_{v}f_{s} =0 \,,
\end{align}
assuming that the plasma in absence of the test charge is field free
and uniform. Furthermore, we assume that the perturbation is week, and
the distribution can be linearized:
\begin{align}\label{pv1}
  f_{s} &= f_{s,0}(\vec{v}) +
  f_{s,1}(\vec{x},\vec{v},t)\\\nonumber
  \frac{\partial f_{s,1}}{\partial t} + (\vec{v}\cdot\vec{\nabla})f_{s,1} &= \frac{q_{s}}{m_{s}}(\vec{\nabla}\Phi)\cdot \vec{\nabla}_{v}f_{s}\\\nonumber
 \sum\limits_{s} n_{s,0}q_{s}\int f_{s,0} \md^{3}v & =0 \,.
\end{align}

The Fourier-Laplace-transform (and the back-transform) of the 
perturbed Vlasov equation Eq.~(\ref{pv1}) are given by
\begin{align}\\\nonumber
  \tilde{h}(\vec{k},\vec{v},\omega) &= \int
  h(\vec{r},\vec{v},t)e^{-i(\omega t+\vec{k}\cdot\vec{r})}\md t \md^{3}r\,,\\\nonumber
  h(\vec{r},\vec{v},t) &= \int
  h(\vec{k},\vec{v},\omega)e^{i(\omega t+\vec{k}\cdot\vec{r})}\md\omega\md^{3}k\,.
\end{align}
We further assume that the
test particle is at rest at $\vec{r}=\vec{0}$, and find:
\begin{align}
 - i \vec{k}\cdot\vec{v} \tilde{f}_{s,1}= -i
  \frac{q_{s}\Phi}{m_{s}}(\vec{k}\cdot \vec{\nabla}_{v}\tilde{f}_{s,0})\,.
\end{align}
For isotropic distribution functions we can change to spherical
coordinates, and assuming that the integral does not depend on $\dphi$ and
$\dtheta$,  replacing
$\vec{k}\cdot\vec{v}=k v\cos\xi$, the Fourier-transform then
simplifies to:
\begin{align}
  -i k v \cos\xi \tilde{f}_{s,1}=- i k \cos\xi \frac{v}{v}
  \frac{q_{s}\Phi}{m_{s}}\frac{d}{d v}\tilde{f}_{s,0}\,,
\end{align}
where also the velocity dependence was changed to spherical
coordinates. From the above equation we find, after the back-transformation:
\begin{align}
  f_{s,1} = \frac{q_{s}\Phi}{m_{s}} \frac{1}{v}\frac{d}{d v}f_{s,0}\,.
\end{align}
Inserting in Eq.~(\ref{p1}) and solving the equation we find
\begin{align}
  \nabla^{2}\Phi - \left[-\sum\limits_{s}\frac{
  n_{s}q_{s}^{2}}{\epsilon  m_{s}}\int\frac{1}{v}\frac{d}{d v}f_{s,0}\md^{3}v\right]\Phi
  = \frac{1}{\epsilon}\delta({\vec{r}})\,,
\end{align}
where the Debye length for collisionless distribution functions is
\begin{align}
  \frac{1}{\Lambda_{D}^{2}} &= -\sum\limits_{s}\frac{
  n_{s}q_{s}^{2}}{\epsilon m_{s}}\int\frac{1}{v}\frac{d}{d v}f_{s,0}\md^{3}v = \sum\limits_{s}\Lambda_{D,s}^{-2}\\
\end{align}
and finding that
\begin{align}\label{nled}
  \frac{1}{\Lambda_{D}^{2}} = -
  \sum\limits_{s}\omega_{p,s}^{2}\int\frac{1}{v}\frac{\partial
  f_{0s}}{\partial v}\md^{3}v\,.
\end{align}

We stay here with the name Debye length rather than using shielding length.
We also point out that this Debye length is only valid for test charges
at rest. For moving test charges the shielding is also discussed by
\citep[][chapter
11.1]{Krall-Trivelpiece-1973}. The term $\omega_{p,s}^{2}=\frac{4\pi n_{s}e^{2}}{m_{s}}$
is the plasma frequency of species $s$.

Inserting $f_{0,s}=\tilde{f}$ and dropping the index $s$ we have
\begin{align}\nonumber
  \Lambda^{-2} &= \omega_{p}
                 \frac{2\pi}{\sqrt{\pi^{3}}\eta^{\thalf}\Theta^{3}U\left(\thalf,\frac{5}{2}-\zeta.\eta\xi\right)}\\\nonumber
  &\qquad\left(
  \frac{2\zeta}{\eta\Theta^{2}}\int\limits_{0}^{\infty}
  \left[\left(1 +
  \frac{v^{2}}{\eta\Theta^{2}}\right)^{-\zeta-1}e^{-\xi \frac{v^{2}}{\Theta^{2}}} \right]
    v^{2} \md v \right.\\\nonumber
 &\qquad\left. + 
    \frac{2\xi}{\Theta^{2}}\int\limits_{0}^{\infty}
    \left[\left(1 
  \frac{v^{2}}{\eta\Theta^{2}}\right)^{-\zeta}e^{-\xi \frac{v^{2}}{\Theta^{2}}} \right]
  v^{2} \md v
  \right)\\\nonumber
&= 2\Lambda_{M}^{-2} \left(
  \frac{\zeta}{\eta}\,\frac{U\left(\thalf,\thalf-\zeta,\eta\xi\right)}
  {U\left(\thalf,\frac{5}{2}-\zeta,\eta\xi\right)} +
  \xi 
  \right)\,.
\end{align}
The additional factor comes from our definition of the Maxwellian.

\bsp
\label{lastpage}

\end{document}